# Polarized pulse pair observations during a long duration interstellar communication experiment

*William J. Crilly Jr.*

Green Bank Observatory, West Virginia, USA

***Abstract*—In prior work, conducted since 2017, two celestial pointing directions have been observed to be associated with the measurement of anomalous high counts of narrow bandwidth, short duration, polarized radio frequency pulse pairs. The prior experimental work utilized up to three geographically-spaced synchronized radio telescopes, a single-dish radio telescope, and a radio interferometer. The experimental work reported here examines full right ascension coverage at one declination, utilizing the interferometer, during 124.1 days. Results suggest the possible presence of an additional anomalous celestial pointing direction. Seven standard deviations of noise-modeled shifts of mean polarized pulse pair count were observed in three celestial directions. Indications of interferometer space delay aliasing were observed. A phase noise test, celestial source identification methods and associated measurements were used to seek potential explanations of the unusual observed phenomena.**

## I. Introduction

### Prior work

Unusually high counts of unexplained narrow bandwidth polarized pulse pairs have been observed in previous work **[1][2][3][4][5][6][7][8]**, conducted since 2017. Three different radio telescope experimental methods have been used, i.e. a geographically spaced synchronized radio telescope system, a single-dish radio telescope, and a two-element radio interferometer. A celestial direction of interest (DOI) was initially found in 2018-2019 observations near 5.25 hr ± 0.15 hr right ascension (RA), -7.6º ± 1º declination (DEC), having anomalous narrow bandwidth pulse pair counts, using synchronized geographically spaced radio telescopes **[1].** The first DOI was re-observed, and a second DOI was identified near 8.84 hr RA, -4.3º DEC in 2024, during interferometer observations. **[5][6][7][8]** An additive white Gaussian noise (AWGN) model has been used to predict residual pulse pair likelihood.

### Objectives

The primary overall objective of this experimental work is to refine, test and provide measurements to help the analysis of possible hypotheses that might explain narrow bandwidth polarized pulse pairs, e.g. hypotheses involving radio frequency interference (RFI), equipment and algorithm issues, natural celestial objects, and communication transmitters of extraterrestrial origin.

The radio telescope experimental systems have been designed to seek energy-efficient, high information rate communication signals having properties indicating that the signals are designed to be readily discovered. Properties include long term repetition of two-tone coherent pulses, i.e. pulse pairs, that are thought to provide measurements unlikely to arise due to natural object emissions. Pulse pair transmissions are conjectured to be possible signatures of celestial communication signals, i.e. detectable artifacts of high information capacity communication channels. **[1]**

The objective of this report is to describe pulse pair observations during a 124.1 day experiment using the radio interferometer, reporting evidence and analysis of the full RA range at one DEC.

### Directions of interest

A DOI was identified in prior work, initially based on two-telescope synchronized measurement of four high signal to noise ratio (SNR) pulse pairs, among ten highest SNR pulse pairs. The DOI pulse pairs were observed between 5.156062 and 5.270635 hr RA, at -7.6º DEC, using the Forty Foot Educational Telescope at the Green Bank Observatory time-synchronized with the sixty foot diameter Paul Plishner Telescope of the Deep Space Exploration Society near Haswell, Colorado. The ten highest SNR pulse pairs were observed after processing telescope data from autonomously recorded transit scans during four days.**[1]**

Further anomalies at this DOI were observed in subsequent work.**[2][3][4][5][6][7][8]** A second DOI was identified during the second-level post-processing of two element interferometer observations.**[7][8]**

Two additional DOIs are suggested in the interferometer observations reported in this work, one conditionally identified as possible RFI, or natural emissions that have properties similar to that of RFI, due to its pulse pair phase properties and associated measurements. The signals observed from the other three DOIs have an unknown cause.

The remaining sections of this report describe a proposed hypothesis, method of measurement, categorized observations, discussion, conclusions and further work.

## II. Hypothesis

### Hypothesis

Radio interferometer measurements of hypothetical narrow bandwidth pulse pairs, conjectured to be due to one or more unknown celestial phenomena, are predicted to indicate high values of pulse pair counts in concentrated celestial directions, during long duration experiments that use a relative direction of arrival measurement system. Falsification of the hypothesis occurs if pulse pair count effect size is low, as RA populations are compared.

## III. Method of Measurement

### Summary

The interferometer system is described in **Fig. 1** and in previous reports **[5][6][7][8]**, and is summarized as follows. Two ten foot diameter offset-fed paraboloid antennas are placed on an East-West baseline, 33.0 wavelengths apart, transit recording at -4.3º DEC celestial direction. GPS satellite-disciplined frequency and time signals are used in an RF to in-phase/quadrature (IQ) baseband downconverter, and multichannel digitizers are used to provide signals applied to first level processing. First level processed files contain measurements of 3.7 Hz bandwidth component pulses in pairs, while each component pulse in a pulse pair exceeds an 8.5 dB SNR threshold at each antenna element.

William J. (Skip) Crilly Jr. is a Volunteer Science Ambassador in Education & Public Outreach of the Green Bank Observatory.
email: wcrilly@nrao.edu

Complex visibility, $V(\mathrm{MJD}, \tau_{\mathrm{INT}})$ is measured using a 50 MHz bandwidth FX correlator **[9],** where MJD is derived from UTC, quantized to 0.25s indicated intervals, and $\tau_{\mathrm{INT}}$ is the interferometer instrument setting used to compensate for instrument antenna element differential signal delay. Second-level processing algorithms examine narrowband signal phase measurements to seek pulses in 0.27s duration pulse pairs that indicate a low difference in component pulse directions of arrival, quantized to a 0.0075 hr RA bin width. Statistical effect size calculation of pulse pair count per RA bin is performed using effect size Cohen's **d**, **[10]** based on a binomial event probability model, expected if measured pulse pair events are due to AWGN. Associated measurements, e.g. frequency spacing of simultaneous pulses and simultaneous power measurements, are performed in first and second level processing, to help in the understanding of the phenomena. Measurement duty cycle is 1/6, providing two 0.27s adjacent-time measurements per 3.0s digitizer trigger period. The polarization of signal measurements is right-hand circular, in the IEEE (1997) definition.**[9]**

## Measurements

The measurement results reported here generally use previous interferometer methods, antenna and measuring receiver settings, and presentation of results, **[5][6][7][8]**, with the following exceptions and details:

**Full RA coverage:** The presentation of results covers 24 hours of RA range, over MJD ranges of 124.1 days, partially overlapping the MJD and RA ranges in previous reports.**[7][8]** The recent previous report processed an RA range of 5.1 to 9.1 hr RA, within a noise-model peak event probability.**[8]** Increased RA range in the current report aids in a search for events that might help explain unusual observed phenomena. RA bins are center-valued here; past work used bin-edge RA.

**RFI look-forward method:** The evolving RFI measurement, reporting and excision algorithm **[5][6][7][8]** has a mode to provide for a look-forward in time capability, in which spectral segments exceeding the first-level 954 Hz band segment pulse concentration per four hours are tagged as RFI positive band segments, occurring up to four hours later in time than the pulse pair events counted in the same spectral segment. The setting was not enabled in the previous report **[8]** and is enabled in this report. The look-forward method raises the possibility of reporting false RFI positives, because some of the 954 Hz spectral segments in the 50 MHz range naturally exceed the RFI pulse pair count threshold, due to AWGN-cause pulse pairs accumulated in spectral segments during four hours.

**Phase noise test signal:** A system test using a uniform-distributed $[-\pi,\pi]$ RF phase angle of the east interferometer element signal has been added, to simulate the response of the measurement system to random phase noise modulating the east interferometer element signal. Applying an artificial phase noise signal is useful to seek natural object, RFI, software and algorithm-induced false positives. A presence of anomalous responses, when the phase noise modulation is enabled, using multiple random value seeds, adds support for a non-celestial auxiliary hypothesis, leading to investigation that might explain measured phenomena.

**Isolated Cohen's d false positive outliers:** Cohen's **d** measurements of pulse pair count, at the top of the increasing $|\Delta_{\Delta f}\Delta_{EW}\phi|$ sorted heap, are naturally high due to the low likelihood of an RA bin having a pulse pair at the top of a relatively large sorted heap. These sporadic outliers were not excised.

**Cohen's d concentration correlations:** Cohen's d measurements within an RA bin are not independent of each other and cannot be directly summed to indicate overall effect size. Rather, they indicate the degree to which an RA bin contains sustained anomalies in the sorted heap at the point of the candidate event in the heap. The narrow setting of the $|\Delta_{EW}\phi| < 0.10$ radian filter is expected to contribute to a somewhat random placement of candidate pulse pairs in the filtered sorted heap. Adjacent RA bin high values of Cohen's **d** may contribute to the effect significance.

**Excision of Sun transits:** Pulse pair measurements during transits of the Sun were excised using a calculated RA and MJD filter, due to highly correlated SNR outliers adding many false positives to the pulse pair candidate collection.

**Determination of directions of interest:** DOIs are defined to be concentrations of pulse pairs observed in a limited RA range. The DOIs are manually chosen by seeking concentrations of Cohen's **d** values consistently above a value of three standard divisions, within one or more adjacent or aliased RA bins. Associated measurements of DOIs are studied to try to find causes of the anomalies.

**Cohen's d details:** The Cohen's **d** measurement calculates the effect size of pulse pair counts in a filtered range of differential phase measurements that result from a modeled near-equal arrival angle of simultaneous narrow bandwidth pulses in each pulse pair, detailed in **Fig. 1**. It is expected that if the simultaneous pulses are components of a continuously transmitted sequence of pulse pairs, and the pulses in the pair measure close to the same angle of arrival, within a given RA bin, then an excess count of pulse pairs should appear across the entire filtered range of the $\Delta_{\Delta f}\Delta_{EW}\phi$ sorted heap. Cohen's **d** should then indicate a continuous deviation from zero standard deviations. Alternatively, if excess pulse pairs are only sporadically high in count in the pulse pair heap, and are due to AWGN, one would expect some sporadic high values of Cohen's **d** together with a large number of values in the vicinity of the expected value of Cohen's **d** = 0, in the same RA bin. The significance of a DOI is additionally indicated by high values of Cohen's **d** in adjacent RA bins and in interferometer space-delay aliased RA bins.

**Determination of transmitting characteristics:** An objective of tests and associated measurements of pulse pairs is to provide evidence of the properties of a distant transmitting aperture, whether based on RFI, a natural object, or an intentional interstellar transmitter. Associated measurements include coincident power in 954 Hz and 50 MHz bandwidths, interferometer visibility magnitude, $\log_{10}$ likelihood of pulse pair component frequency spacing, MJD, pulse RF frequency, continuum power, $\log_{10}$ likelihood of SNR, and RFI spectral margin. Tests include noise application and modified filters.

**Modified filter test:** Transmitting characteristics may be sought by conducting tests that modify the phase thresholds during processing of candidate pulse pairs. In this report, the $\Delta_{EW}\phi$ and $\Delta_{\Delta f}\Delta_{EW}\phi$ filter settings were adjusted to seek pulse pairs that do not have direction of arrival measurements indicative of hypothesized intentionally transmitted pulse pair signals. The two filters were set in this test to accept pulse pairs that cover a modified range of phase values, so that non-common arrival direction signals can be detected, while suppressing the detection of celestial signals having common angles of arrival. A conditional identification may then be suggested of scattered RFI or celestial natural emitters, e.g. emitters that are expected to have random pulse pair component phase characteristics.

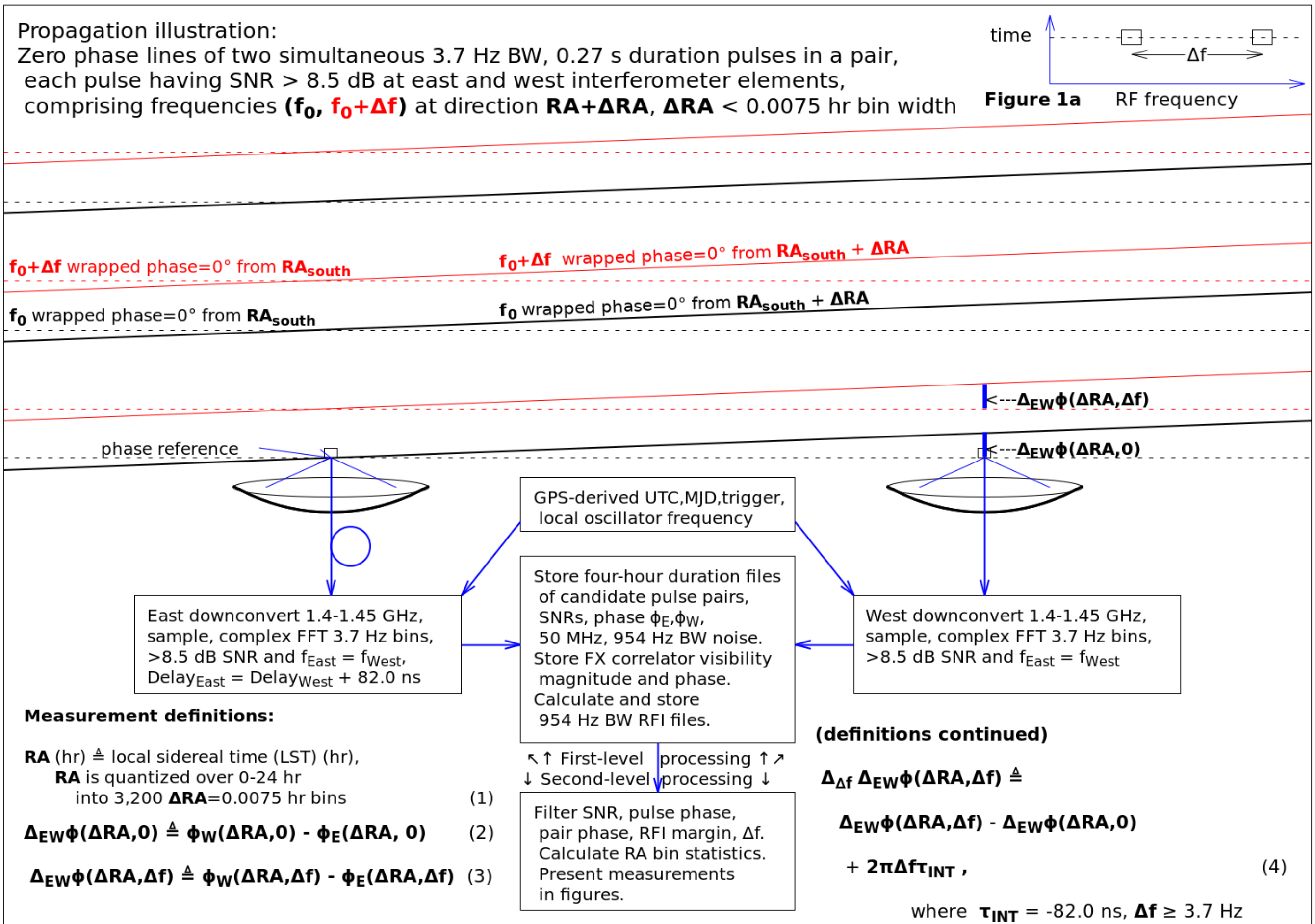

**Figure 1: Signal propagation and processing diagram**
pointing azimuth = 180.0 degrees

**Figure 1:** Measurement definitions (**Equations 1-4),** are used to seek relative angle of arrival within pulse pairs. An illustration of the underlying propagation paths to the interferometer elements is presented. The block diagram summarizes salient machine steps used to process telescope signals to produce the graphics files displayed in this report. The objective of the signal processing is to seek pulse pairs having low differential phase measurements, indicating that the two pulses in each pair have a nearly common angle of arrival. Additional cable delay is added in the east element signal path, to reduce FX correlator zero-delay-valued residual responses, and compensated in second-level processing.**[9]** Associated simultaneous measurements are performed in processing steps to seek and test various explanations of anomalous pulse pair indications.

## IV. Observations

Observation run O9, reported here, comprises the second-level processing of first-level processed files covering 24 hours of RA at -4.3° DEC during a duration of 124.1 days. Instrument settings are listed above each plot area and apply to all the data plots within each figure. Experimental results are plotted in four general categories as follows:

**Observed DOIs**: **Figs. 2-3** show Cohen's **d** effect size of the pulse pair count, and pulse pair count of four identified DOIs, including interferometer aliased RA directions. RA resolution is expanded so that central and aliased directions may be observed. Full RA coverage is in **Figs. 4** and **21.**

**Phase noise and $\tau_{INT}$ = 0 tests: Figs. 5-9** show full RA views of Cohen's **d**, using four differently seeded east element phase noise signals, and offset interferometer delay, i.e. $\tau_{INT}$ = 0 ns. These system tests are presented to assay the relationships between the element differential phase measurements and the pulse pair count effect size Cohen's **d**, across the full RA range.

**Pulse pair associated measurements: Figs.10-18** show associated measurements, simultaneous with pulse pair candidate events, to seek potential wider bandwidth transient phenomena received at the time of the pulse pair events. RFI and equipment issues also may be studied.

**Alternate source coincidences: Fig. 19** shows MJD and RA of pulse pairs together with excised solar region and the location of the Sun and Moon. **Fig. 20** shows the calculated AWGN-model binomial event probability vs. RA, of the pulse pairs in the sorted heap. **Fig. 21** shows the Cohen's **d** of pulse pair counts in 3,200 RA bins during the 124.1 day experiment, together with MJD 60586 continuum and visibility measurements. **Fig. 22** shows the effect size measurement performed with random phase noise applied to the east element signal, and east and west element continuum plot sensitivity reduced, compared to the higher plot sensitivity in **Fig. 21**. **Figs. 23-24** show pulse pair counts over the full 24 hour range, using telescope and random phase of the east interferometer element, respectively. **Fig. 25** provides test results of indications relevant to gleaning signal source characteristics, described in **Modified filter test** in **III. Method of Measurement. Fig. 26** displays high visibility measurements across MJD and RA.

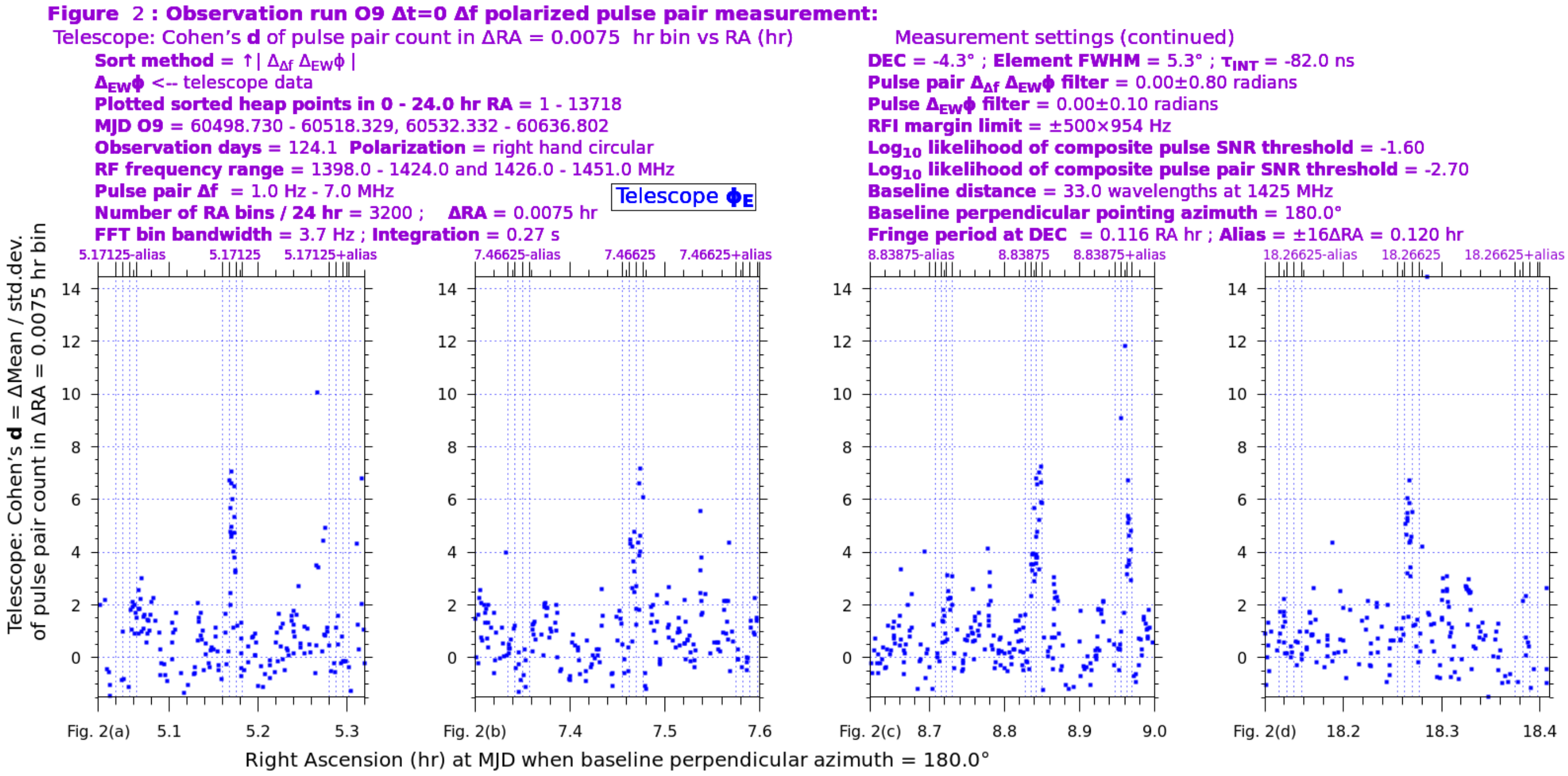

**Figure 2:** The Cohen's **d** effect size of the pulse pair count are plotted at the DOIs together with interferometer-aliased RA bins of a DOI at ±0.120 hr RA. Central and aliased RA indications help in understanding possible celestial causes. The RA values on the upper axis labels the center of the central bin, among three adjacent bins. The apparent sparse presence of DOI pulse pair counts at the expected noise-cause binomial distribution mean count, i.e. near Cohen's **d** = 0, implies that anomalously high mean pulse pair counts are present throughout the 13,718 count sorted heap, covering 124.1 days of observation.

Low $|\Delta_{\Delta f}\Delta_{EW}\boldsymbol{\phi}|$ values are expected to be indicative of Δf frequency-spaced simultaneous pulses having common arrival directions. In the 8.83875 RA pointing direction, aliasing anomalies are apparent in upper and lower geometric delay aliased RA bins. The aliasing is thought to be due to propagation path delay between interferometer elements. Hypothetical received celestial signals are expected to be measurable at an MJD time that resulted in one wavelength of added geometric path delay across antenna elements. The anomalies at 5.17125 hr RA appear to be associated with prior observations. **[1][2][3][4]**

The aliased response in the upper aliased band in **Fig. 2(c)** appears to be highly unusual, given an AWGN explanatory model. Aliasing in the **Fig. 2(c)** lower aliased band at Cohen's **d** ≈ 2-3 is also apparent.

---

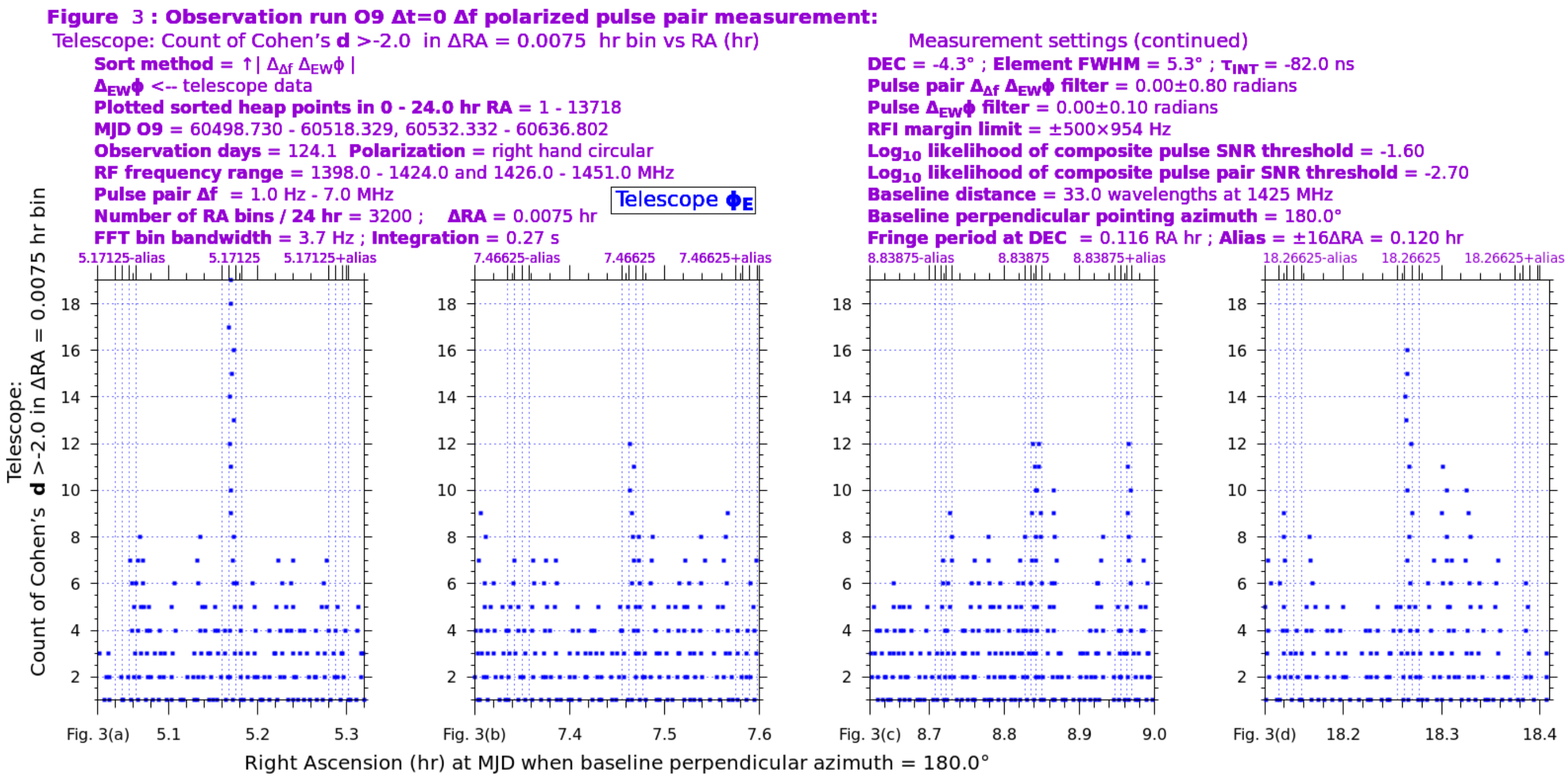

**Figure 3:** The pulse pair count in each RA bin is plotted together with the associated RA measurement. The 8.83875 hr RA events indicate that adjacent RA bins are occupied with pulse pairs. The plotted alias bins are centered at plus and minus sixteen ΔRA bins, rounding the measured 0.116 hr RA alias to 0.120 hr, an integer quantized RA bin.

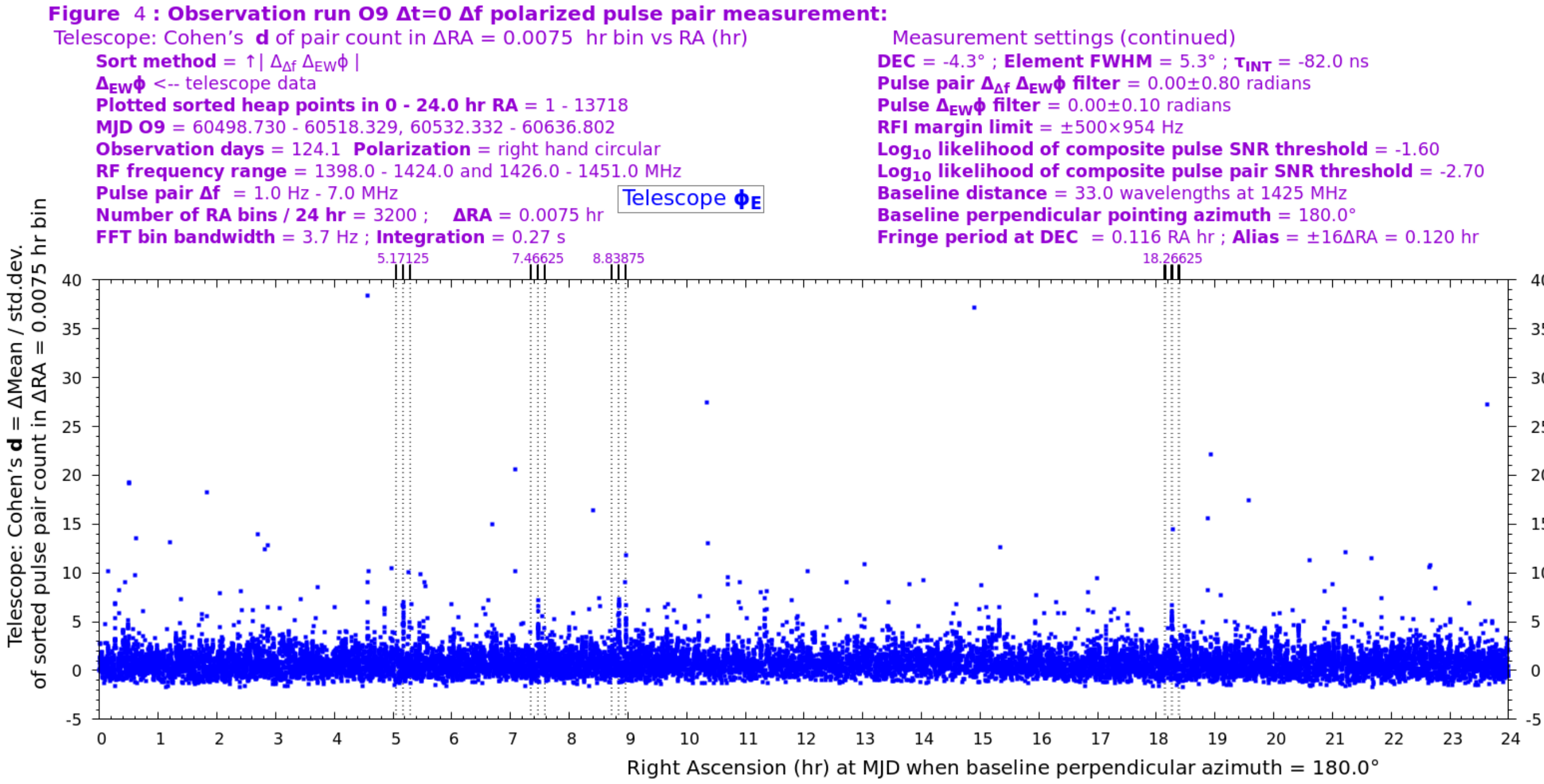

**Figure 4:** The full RA range is displayed to examine the DOI events in comparison to other events. Sporadic high levels of Cohen's **d** are often caused by events at the top of the sorted heap used in the binomial-modeled measurement of Cohen's **d** effect size. Such events are unlikely to be seen in an RA bin due to the large number of RA bins, and therefore indicate high Cohen's **d** values, while the number of binomial trials is low, i.e. at the top of the sorted heap.

---

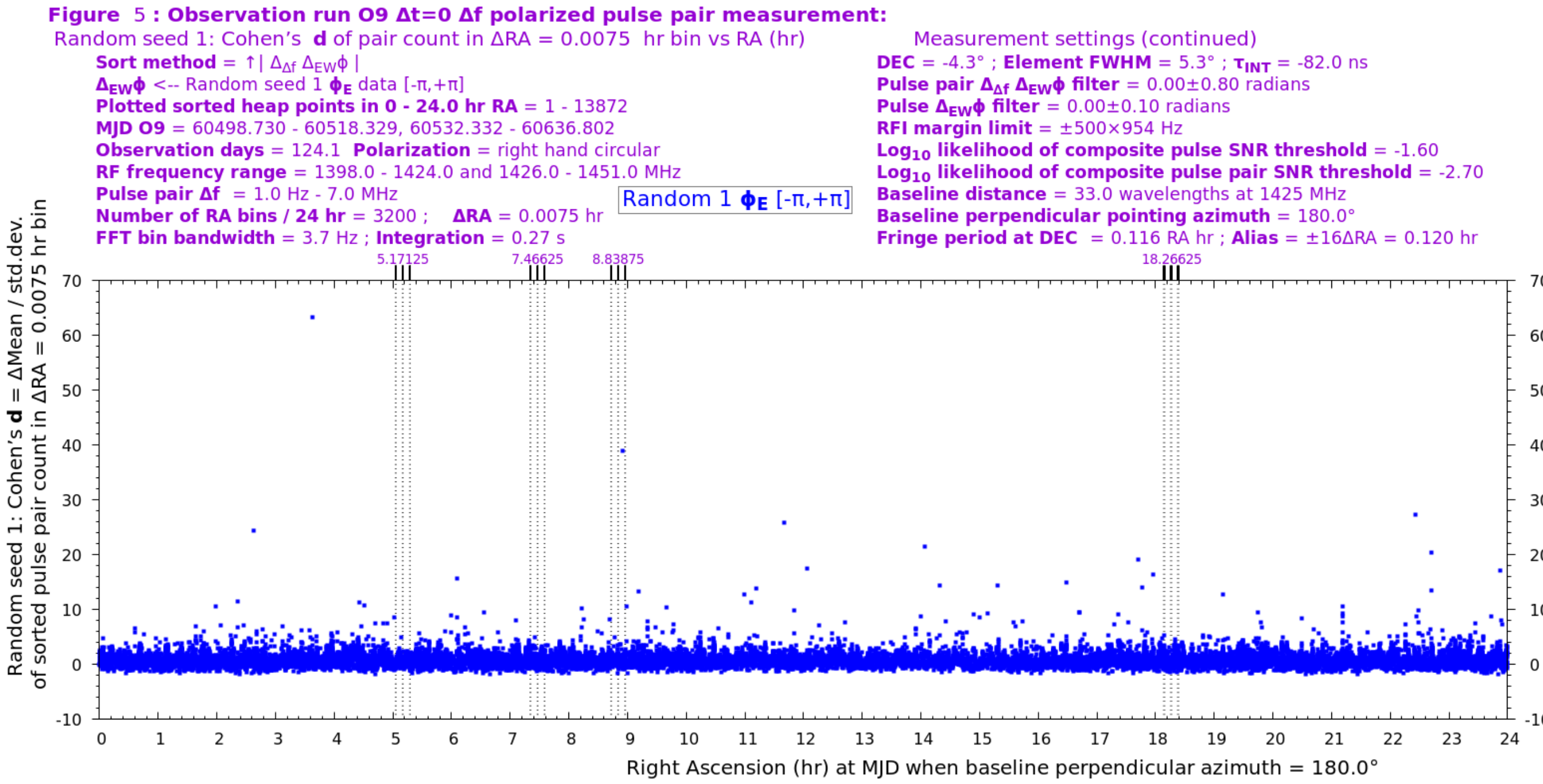

**Figure 5:** A receiver test was performed that artificially changes the value of the east element RF phase measurement to a uniform distributed random value between -$\pi$ and $\pi$. This phase noise setting modifies the phase information in received pulse pairs, required to ascertain a common direction of arrival of the two pulses in the pair. Directions of interest not showing high directionally-filtered pulse pair counts is an indication that the east sky phase measurement contributes to the anomalous counts. First level candidate pulse pairs are selected due to high SNR at both elements. These candidates are filtered in second level processing using a relative angle of arrival measurement system based on phase measurements. **Figs. 6-8** display the Cohen's **d** effect size with incrementing seed of the random number generator that produces the phase noise values. **Fig. 9** shows the measurement with a modified value of instrument delay compensation.

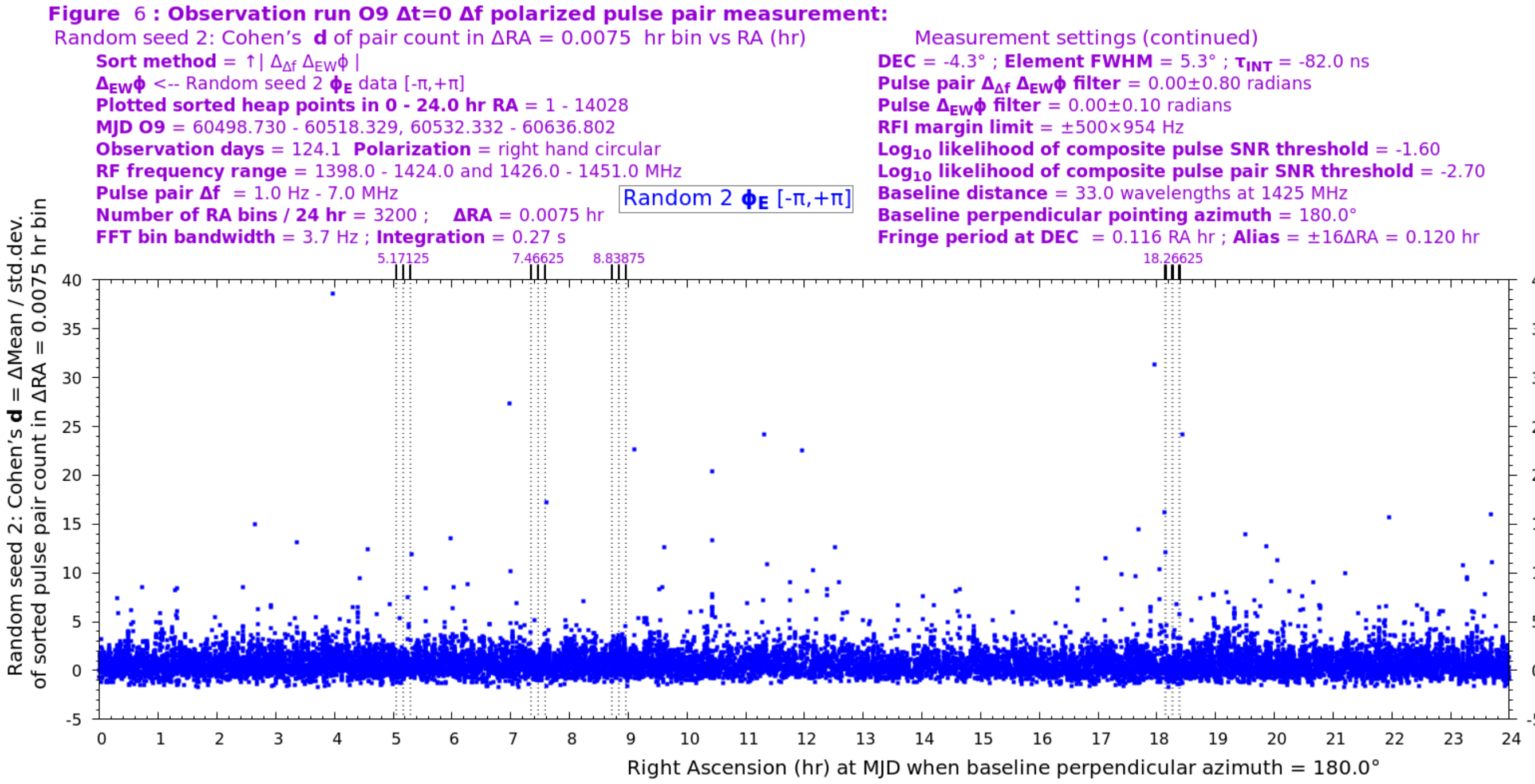

**Figure 6:** The seed for the randomizing function was changed from a value of 1 to 4, incrementally, in **Figs. 5-8.** This random seed change allows the examination of the extent to which artificially produced random phase values in east element measurements produce Cohen's **d** effect size outliers. DOIs are identified in two ways. First, sustained high Cohen's **d** values as the number of pulse pairs per RA bin are counted throughout the second-level filtered and sorted heap. Second, adjacent and aliased RA bins are observed for high values of Cohen's **d**.

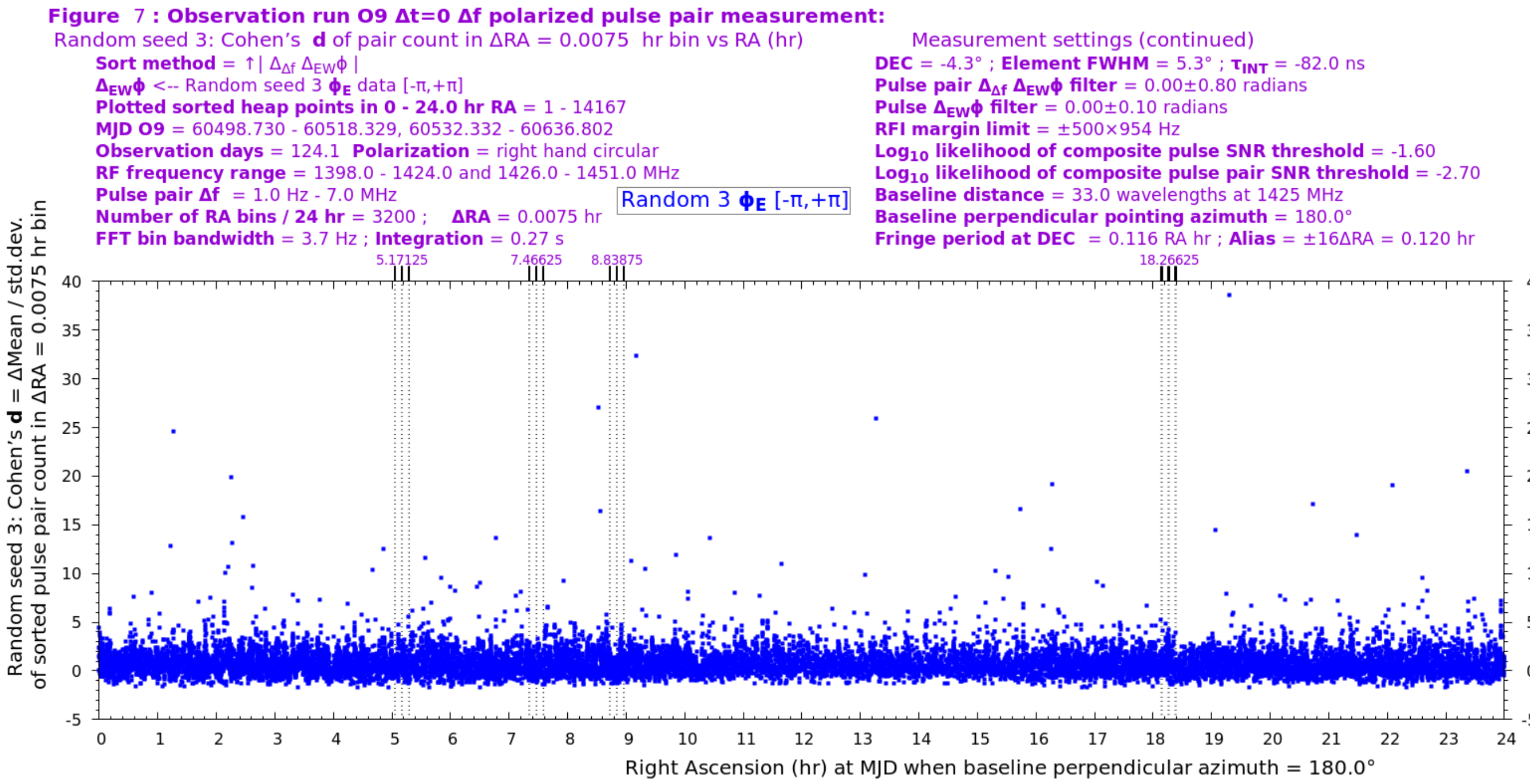

**Figure 7:** The presence of anomalies when random phase is applied may be due to a pre-filter higher pulse pair count in RA directions. For example, if pulse pairs have an anomalous high count in an RA bin, before the $|\Delta_{\Delta f}\Delta_{EW}\phi|$ filter is applied, then it is possible that an anomalous high pulse pair count will similarly occur after the $|\Delta_{\Delta f}\Delta_{EW}\phi|$ is applied to the candidates. This confounding possibility is ameliorated by examining an RA-bin concentration of Cohen's **d** values measured between four and seven binomial distribution standard deviations.

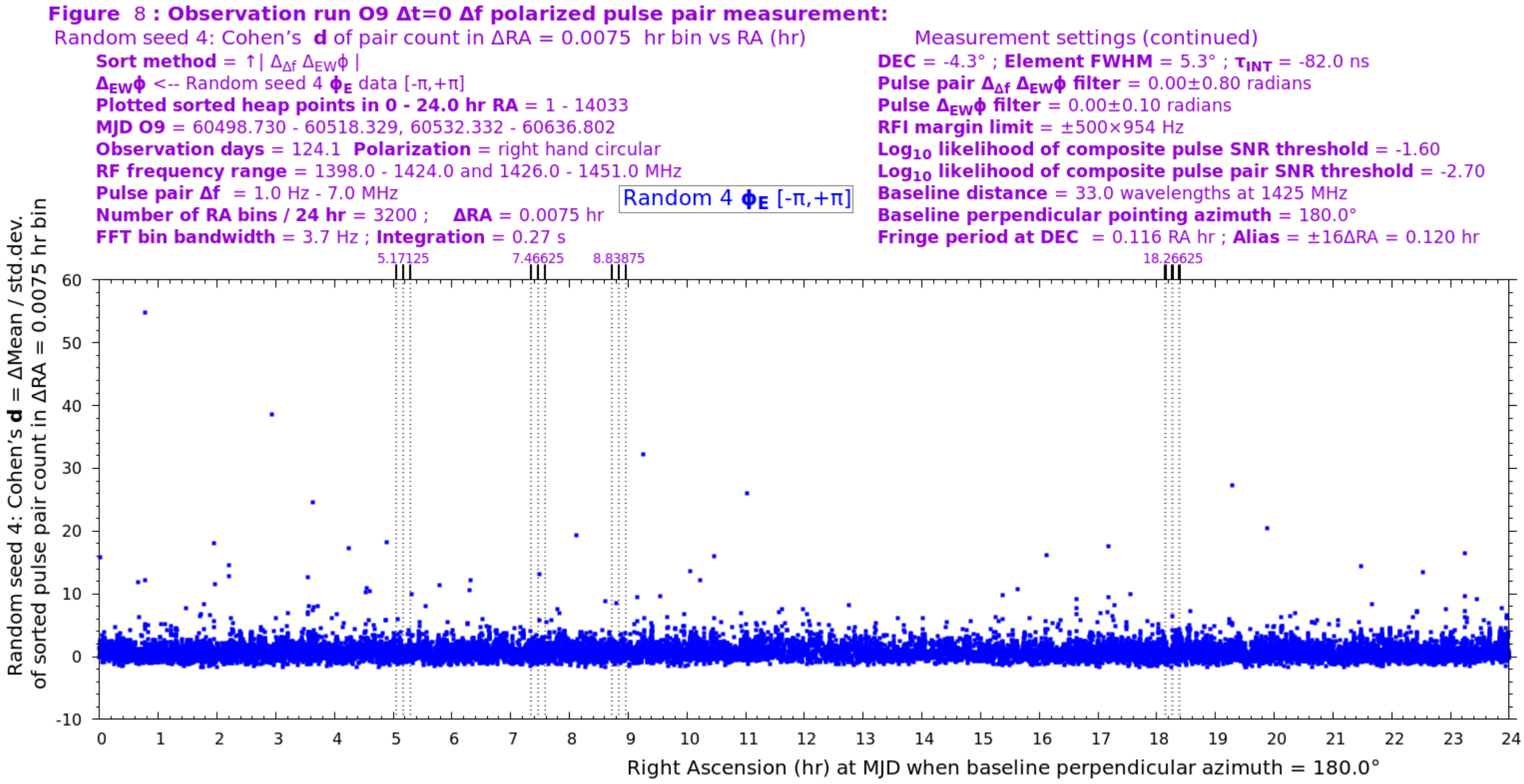

**Figure 8:** The fourth seeded random phase noise application does not indicate RA concentrations of levels of Cohen's **d** in the four to seven standard deviation range. Sporadic isolated high levels of Cohen's d are expected due to pulse pairs located at the top of the filtered and sorted heap.

---

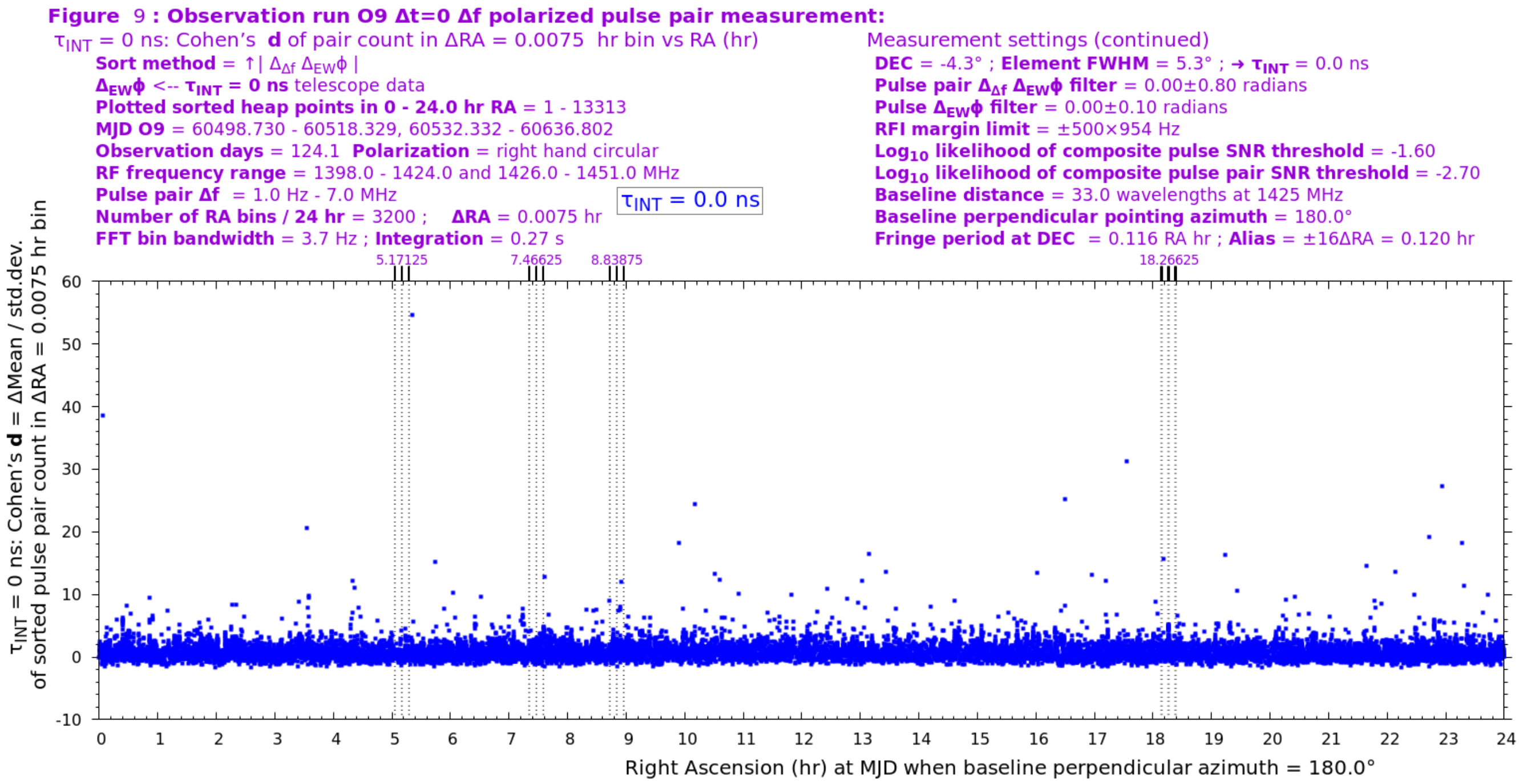

**Figure 9:** Replacement of the interferometer instrument delay $\tau_{INT}$ from -82 ns to 0 ns provides a means to reduce the detection sensitivity of Δf frequency spaced pulses in each pulse pair. Comparison of **Fig. 9** absence and **Fig. 4** presence leads to an inference that the Δf value of frequency-separated, simultaneous pulses in DOI pulse pairs, together with the interferometer delay $\tau_{INT}$, contribute to anomalies in pulse pair count. This observation implies that there is a common direction of arrival of simultaneous narrowband component pulses in each pulse pair, and that the pulse pair frequency spacing is likely an important contributor to the observation of Cohen's **d** anomalies in the effect size measurement across RA populations.

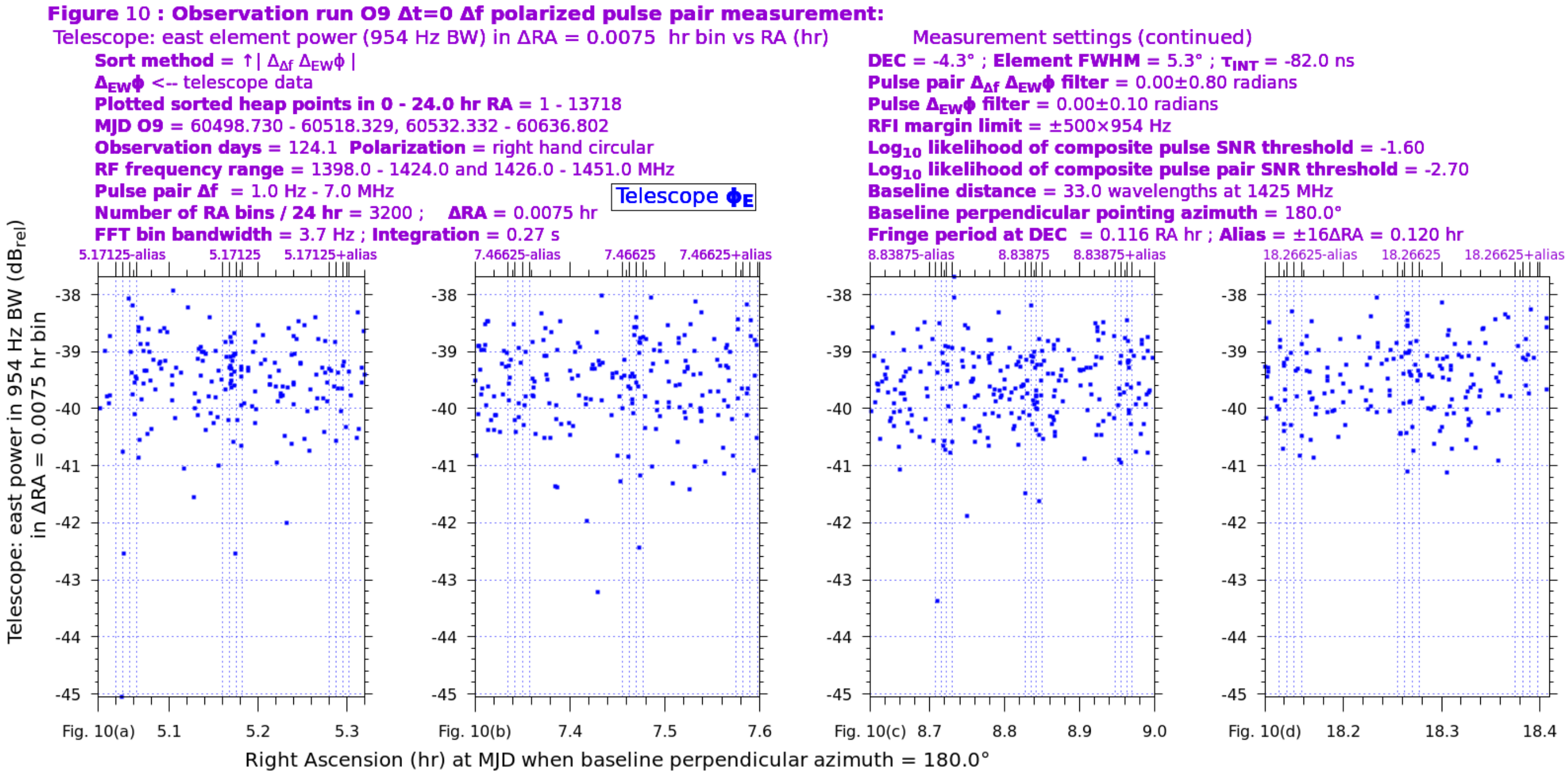

**Figure 10:** Four simultaneous noise measurements are presented in **Figs. 10-13**, to seek RFI and natural object explanations of the anomalies. The east and west interferometer element noise power is measured in 954 Hz (**Figs. 10-11**) and 50 MHz (**Figs. 12-13**) bandwidths, averaged over the 0.27 s FFT time window. In **Fig. 14**, the interferometer 50 MHz bandwidth visibility magnitude is plotted. **Fig. 10(b)** and **Fig. 10(d)** indicate some increase in several measurements of the 954 Hz bandwidth integrated noise power in DOI associated RA directions.

---

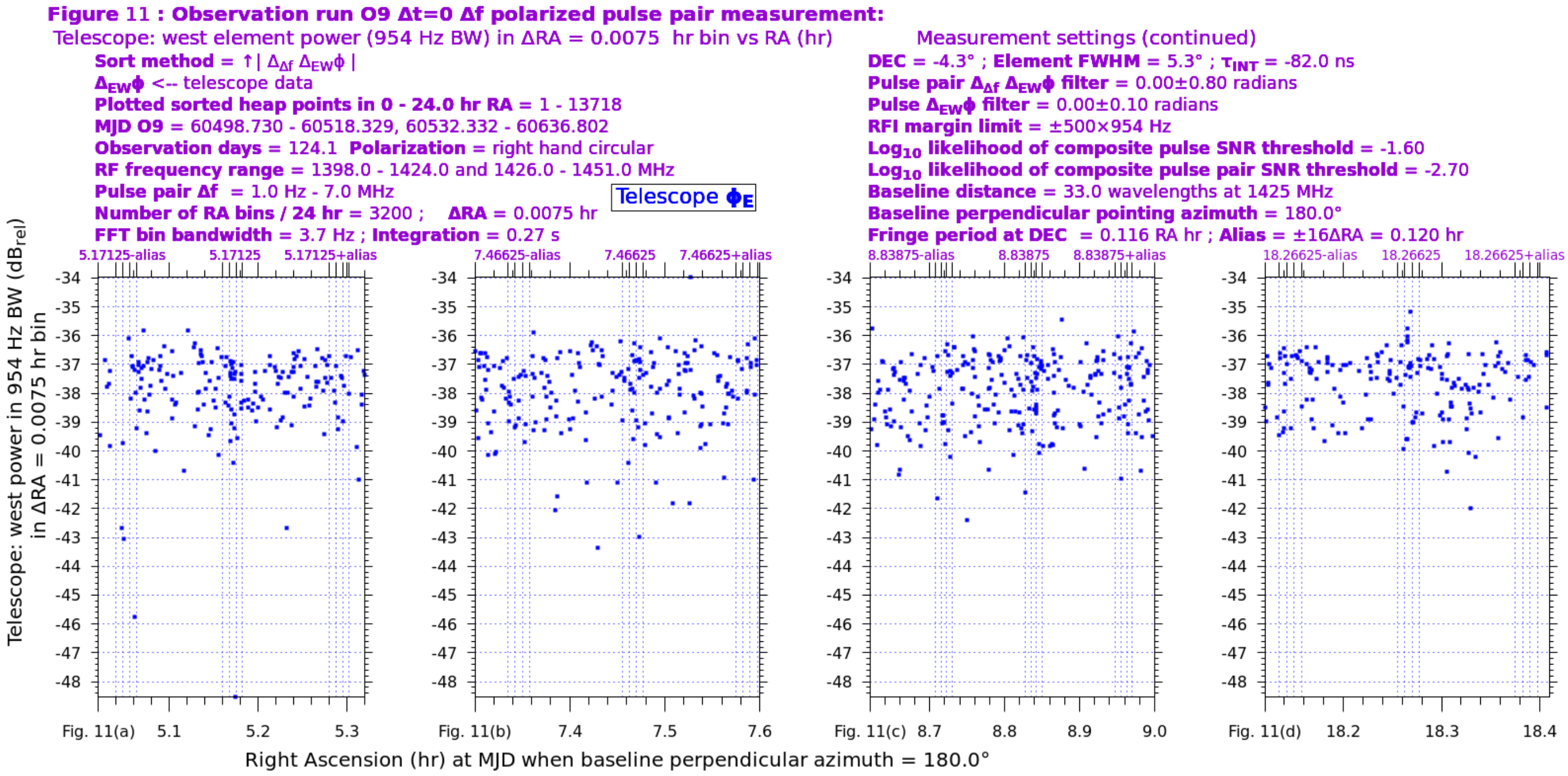

**Figure 11:** West element 954 Hz bandwidth power measurements indicate higher values in **Fig. 11(d)** in the RA bins associated with the fourth DOI, surrounding RA 18.26625 hr. This may imply that other narrow bandwidth signal components are present at the time of a pulse pair event. Another possibility is that the pulse pairs observed in **Fig. 11(d)** are caused by a broadband transmission. **Fig. 13(d)**, displaying 50 MHz bandwidth west element integrated power, seems to contribute evidence to support this idea.

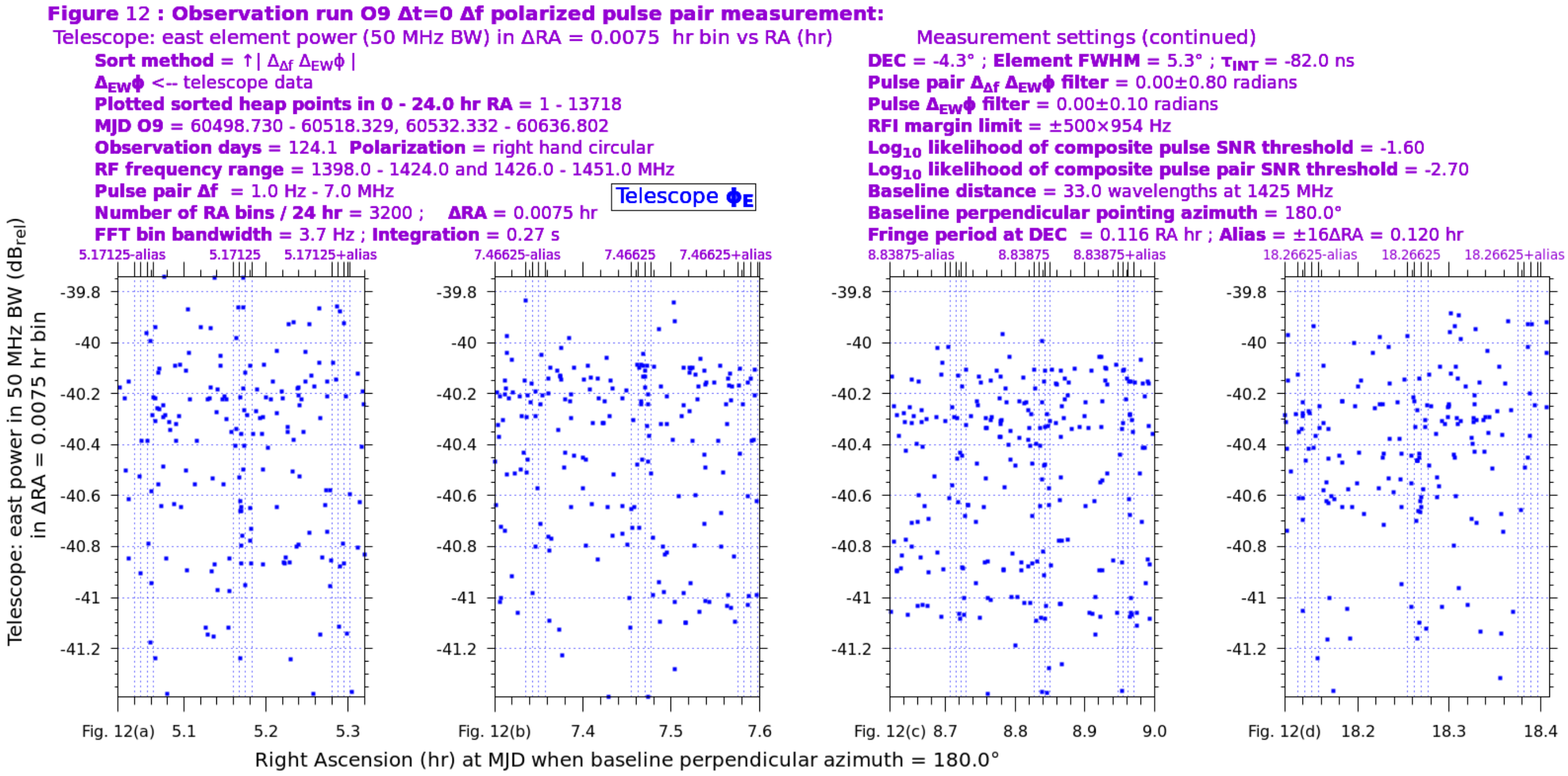

**Figure 12:** Power measurements in a 50 MHz bandwidth are used to examine a possibility that a wide bandwidth event might be causing the high SNR narrow bandwidth pulse pairs. Narrow bandwidth SNR measurements are generally independent of broadband noise power, assuming the noise amplitude statistical properties are unchanged at higher noise power levels. An exception exists in the interferometer due to element-to-element correlated signals and SNR threshold filters with simultaneous SNR measurements, discussed in V. DISCUSSION **Correlation across antenna elements**.

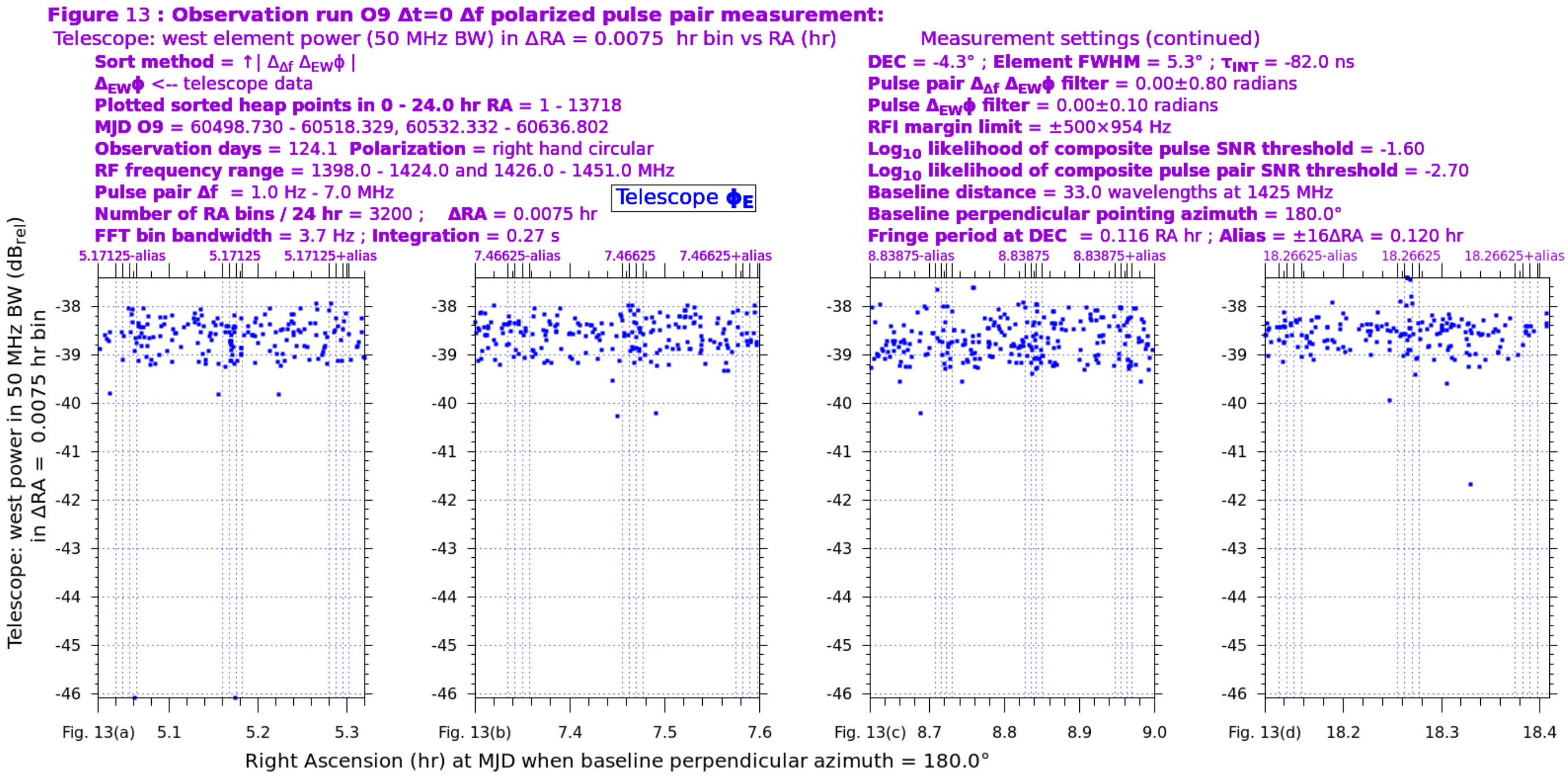

**Figure 13:** A noticeable increase in the west element 50 MHz bandwidth noise power is seen in **Fig. 13(d)**. Differences between the characteristics of this 18.26625 hr RA DOI and other DOIs might be an indication that the cause of the anomalies in this DOI is different than the cause of the anomalies in other DOIs. These differences may also be examined by studying other associated measurements. For example, the absence of high east element 50 MHz bandwidth measurements, in **Fig. 12(d)**, compared to **Fig. 13(d)**, may be an indication that the source of the transmission that resulted in the anomalies in the 18.26625 hr RA DOI is not near the directivity peak direction of each interferometer element. Similarities between **Fig. 10(d)** and **Fig. 11(d)** do not support this idea.

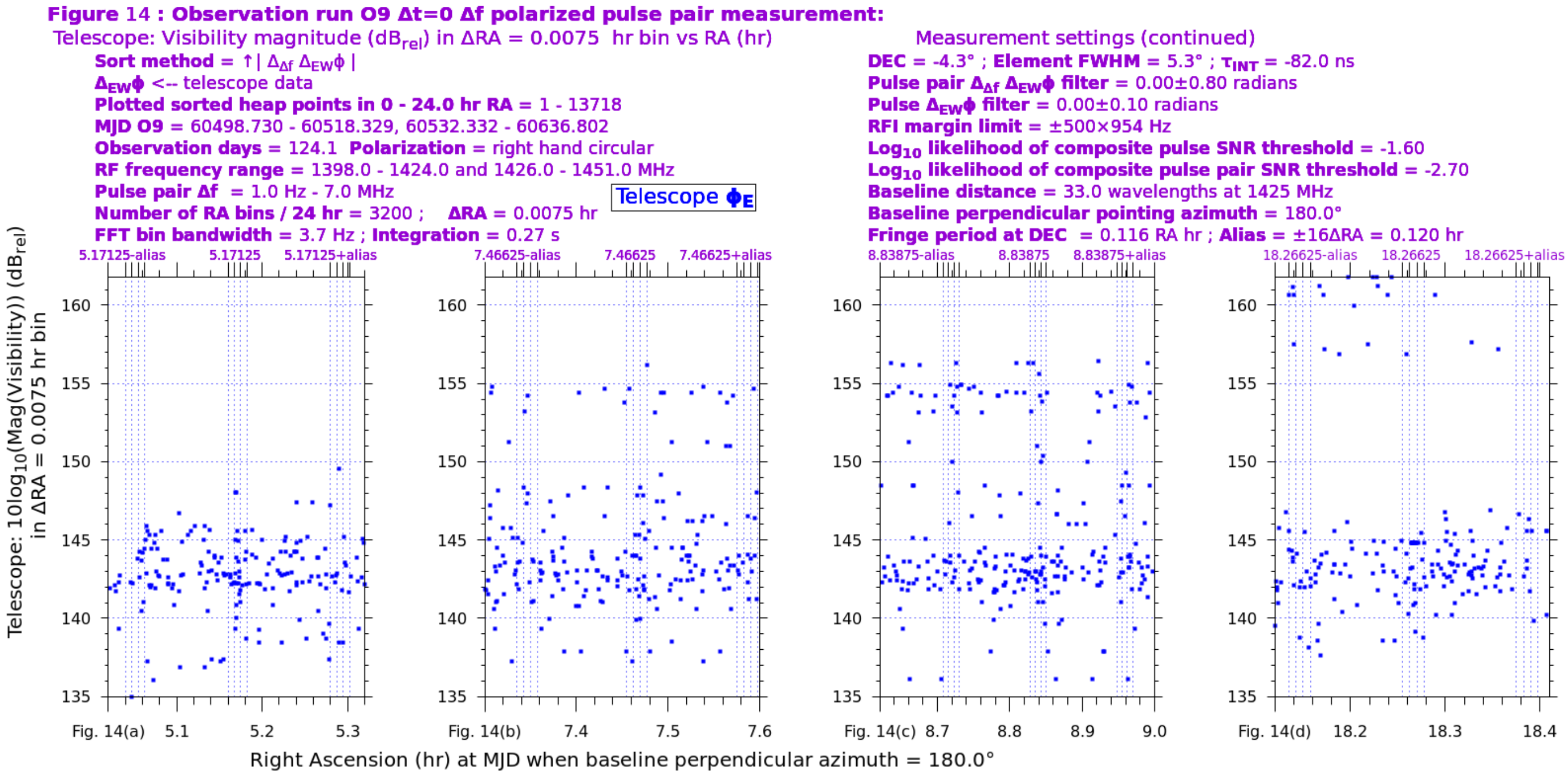

**Figure 14:** A simultaneous 50 MHz bandwidth visibility magnitude measurement provides a way to see if a correlated signal has been propagated to the antenna elements of the interferometer, at the time of the simultaneous narrow bandwidth pulse pair event. A narrow bandwidth pulse pair near the 8.5 dB SNR threshold, by itself, is expected to not have sufficient correlated power to indicate high visibility. Wide bandwidth highly correlated signals may be due to RFI, or a natural astronomical object. Sun visibility responses were excised using an MJD / RA filter, as shown in **Fig. 19**. Full RA and MJD observation of high levels of visibility are plotted in **Fig. 26**.

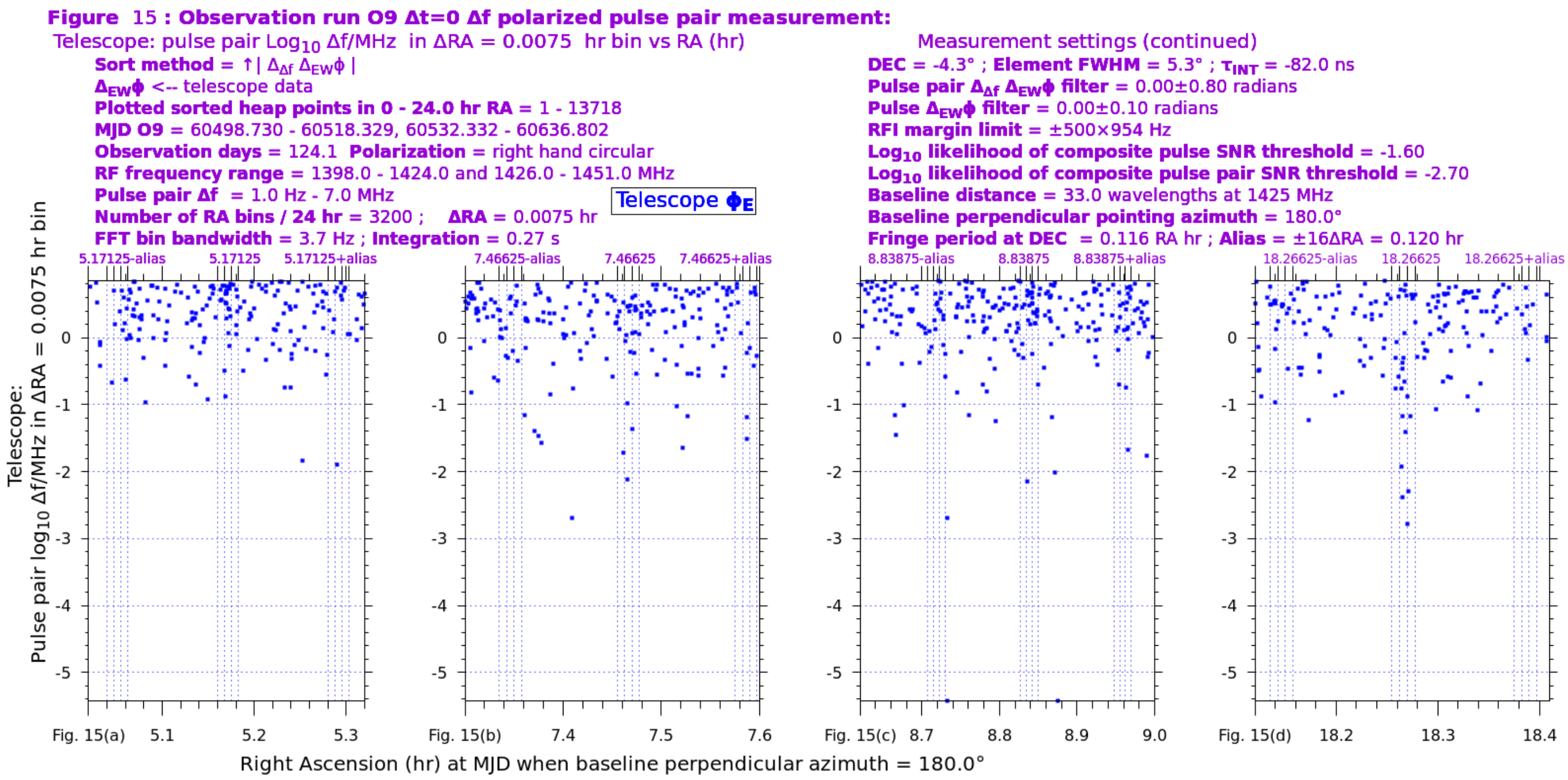

**Figure 15:** The RF frequency spacing, Δf, between the component pulses in each pulse pair was measured and displayed as a logarithmic quantity, to estimate noise-caused event likelihood, given a Poisson distribution model of RF frequency spacing. **[1]**Appendix C. An increased count of pulse pairs in DOI RA bins naturally cause higher counts of low values of Δf, to a degree. $Log_{10}$ likelihood near -3 in **Fig. 15(d)** is significant while maximum pulse pair count of the DOI is 16, shown in **Fig. 3** and **Fig. 23.**

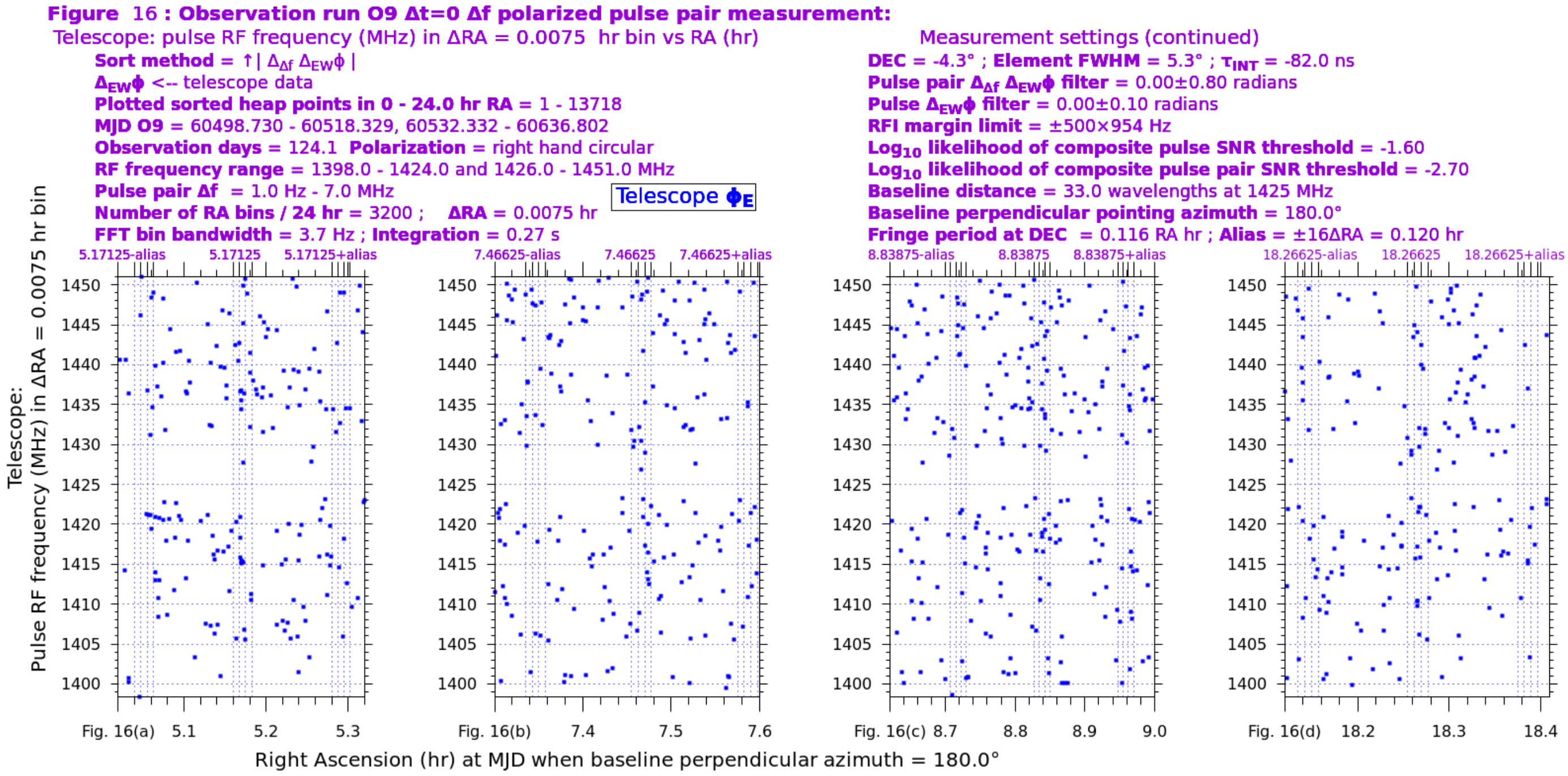

**Figure 16:** The RF frequency of the lower frequency pulse in a pulse pair is displayed for each candidate filtered pulse pair of each DOI. RFI often occurs in frequency bands, and is expected to be reduced in proposed radio astronomy bands. On the other hand, intentionally transmitted signals are expected to occupy widely distributed RF frequencies, while channel capacity is prioritized over signal detectability. **Figs. 16(a-d)** do not indicate clearly apparent concentrations of RF frequency with the exception of ten pulse pairs in the range 5.0-5.1 hr RA and 1420.604-1421.357 MHz, in **Fig. 16(a)**.

---

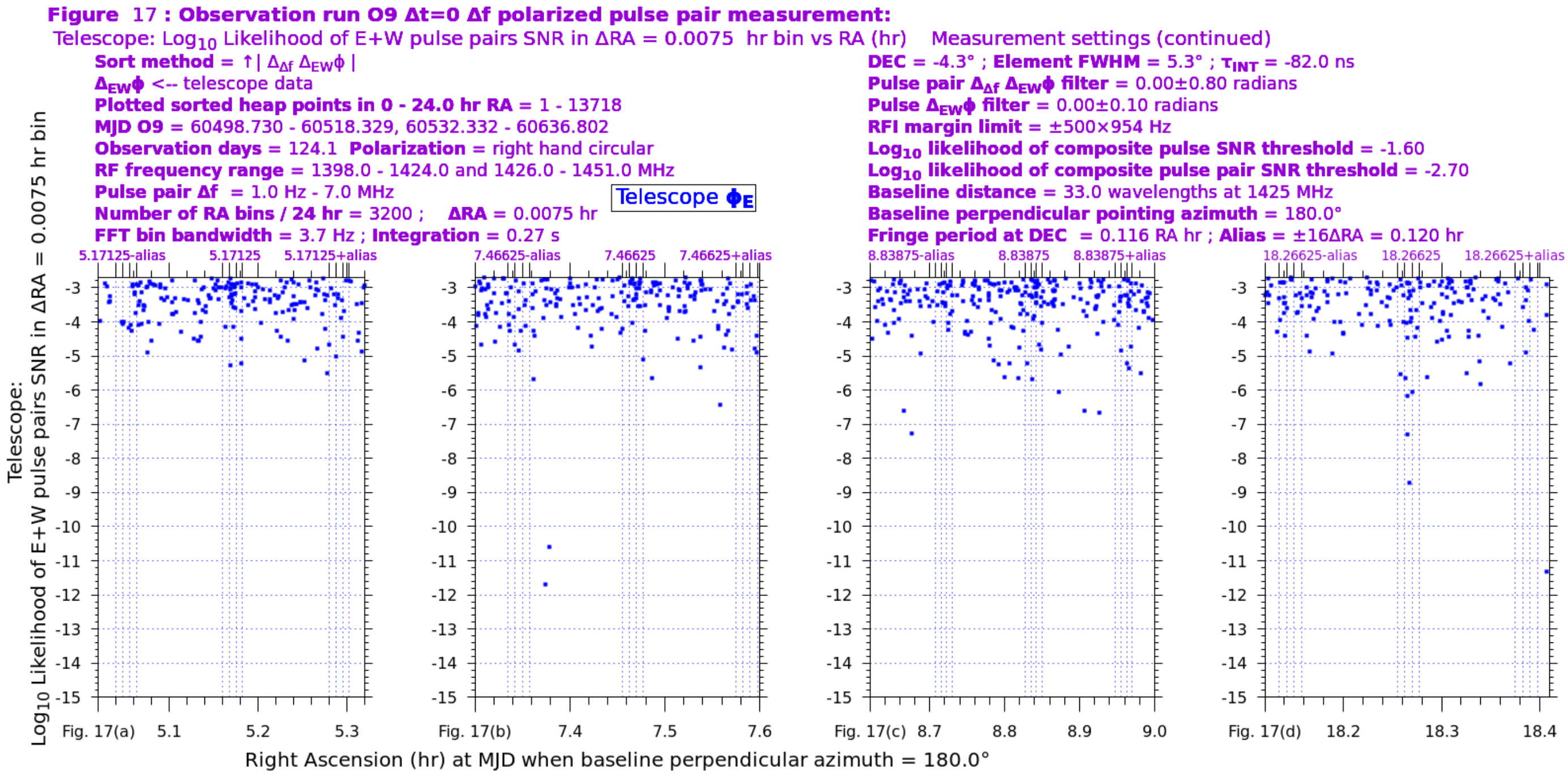

**Figure 17:** SNR is an important measurement to discern intentionally transmitted signals. The importance is not absolute. For example, an intentional transmitter may be designed to limit the transmit power to set a target received power level needed to establish a given channel capacity, detection time and anticipated receiver system, in beamed transmissions. If such a transmitter power-level scenario is commonplace, one might expect to measure similar levels of SNR of candidate pulse pairs, across DOIs.

The SNR of the 18.26625 hr RA DOI, in **Fig. 17(d)**, has a $\log_{10}$ likelihood of SNR significantly lower than the other three DOIs. Together with other associated measurements and tests, high SNR may imply an RFI or natural cause.

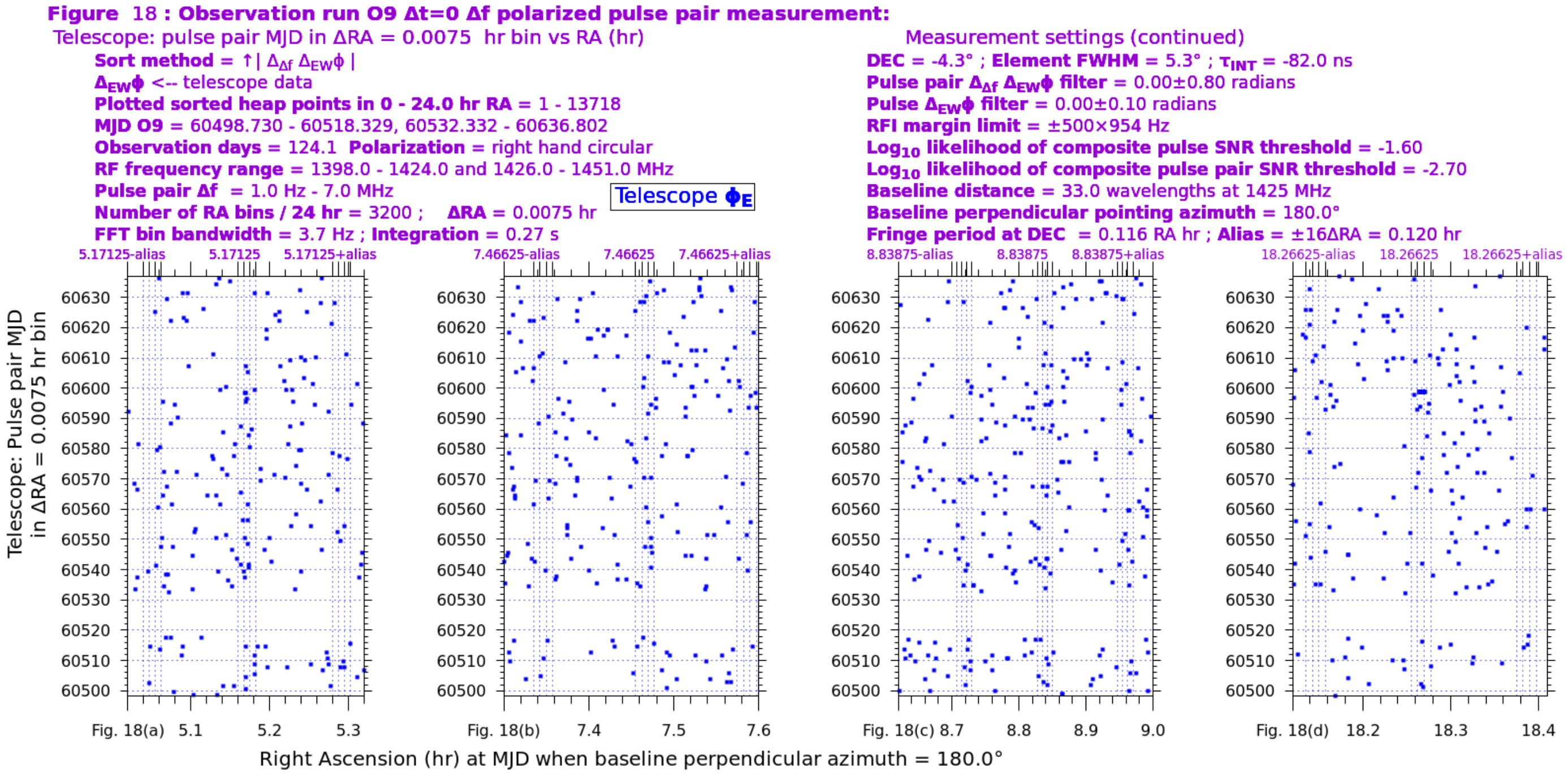

**Figure 18:** Measuring the MJD of a candidate pulse pair is important because some RFI and natural astronomical sources indicate flux variation, concentrated in time. As an example, nine of sixteen pulse pairs in the 18.26625 hr RA bin measure a single-day event time of MJD = 60598.89080 to 60598.89105. This observation, together with other associated measurements may be used in comparisons, across DOIs, to seek characteristics of hypothesized intentional interstellar transmitters, RFI and natural objects.

---

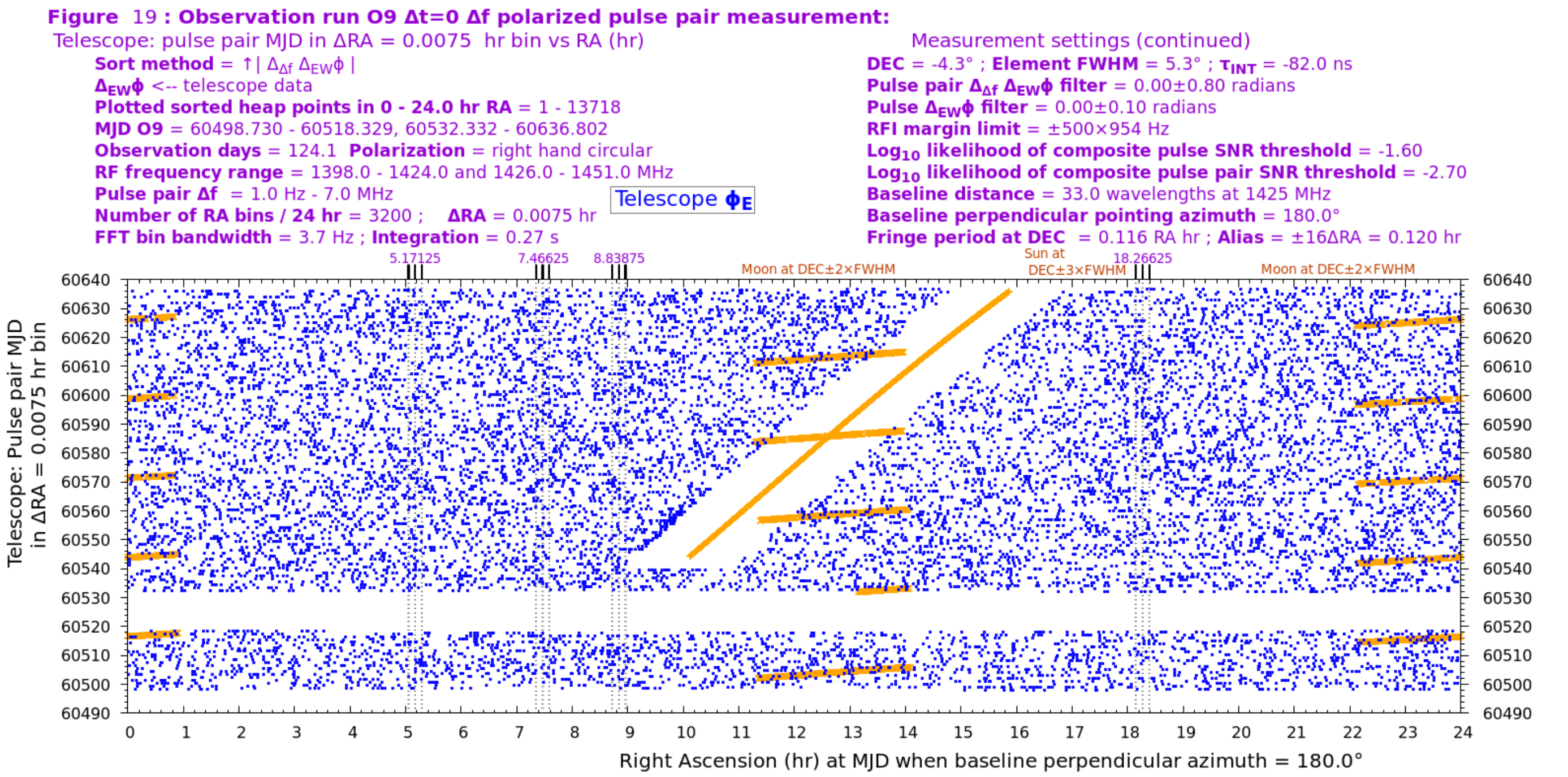

**Figure 19:** Processed pulse pairs are plotted over the full MJD and RA ranges of the experiment. Calculated Sun and Moon coordinates are added, indicating the DEC ranges over which the MJD vs RA lines are placed. The parallelogram-shaped region is due to an algorithmic excision of solar transits while main beam and side-lobe solar power was high, chosen to be ± 1 hr RA pointing and MJD > 60540. A portion of the Sun-caused side-lobe power is visible, indicating that the Sun can contribute to increased pulse pair count. The MJD region gap around 60525 MJD is due to a time period when the radio telescope system was not operating. Pulse pair events during Moon transits were not excised, so that possible anomalous pulse pairs, e.g. RFI reflections from the Moon, may be sought.

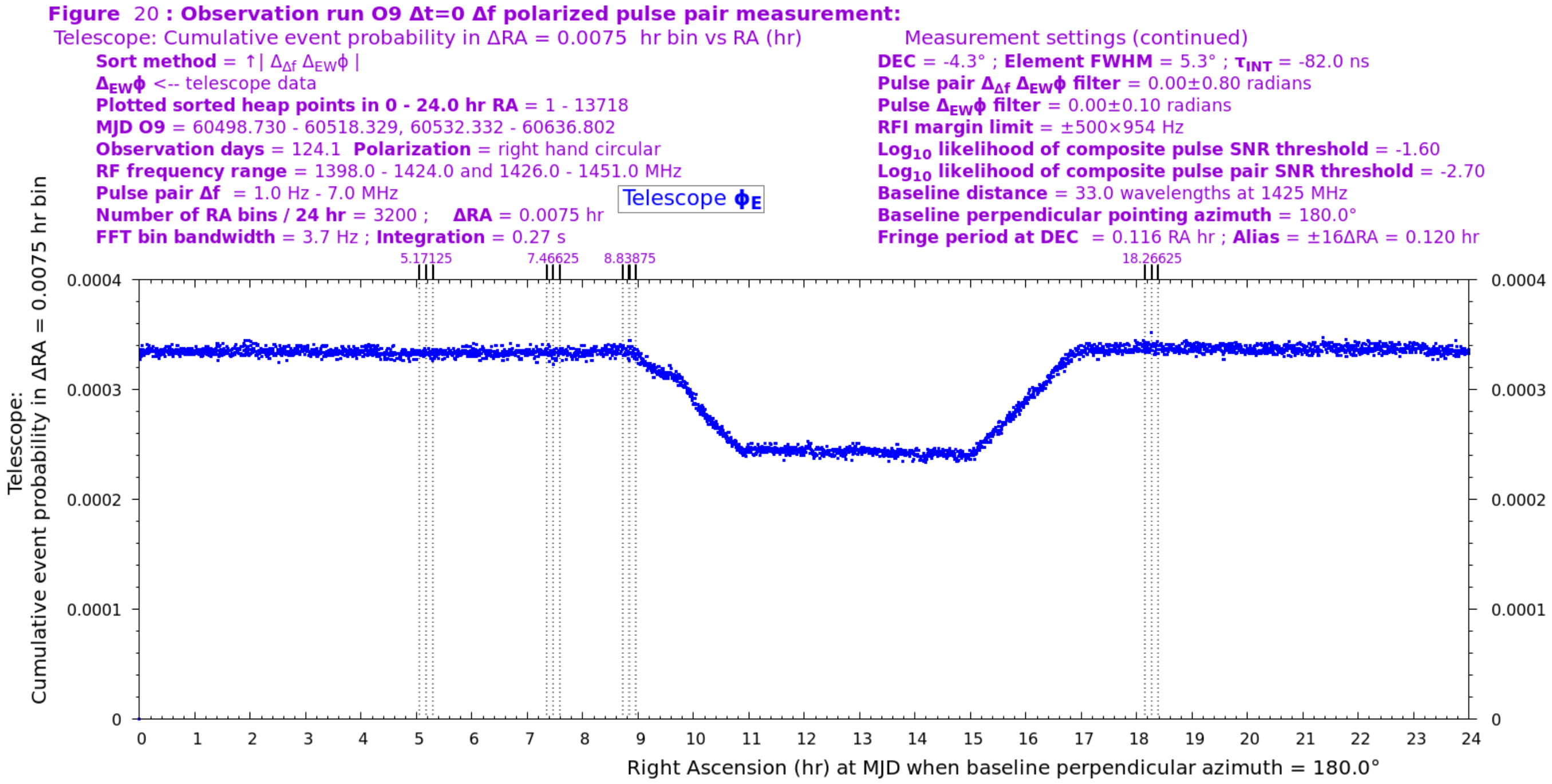

**Figure 20:** An AWGN signal source model applied to a binomial event probability model is used to calculate the probability of a pulse pair event presented in an RA bin. The dip in the event probability is due to the (MJD,RA) region excised due to high Sun noise, shown in **Fig. 19**.

---

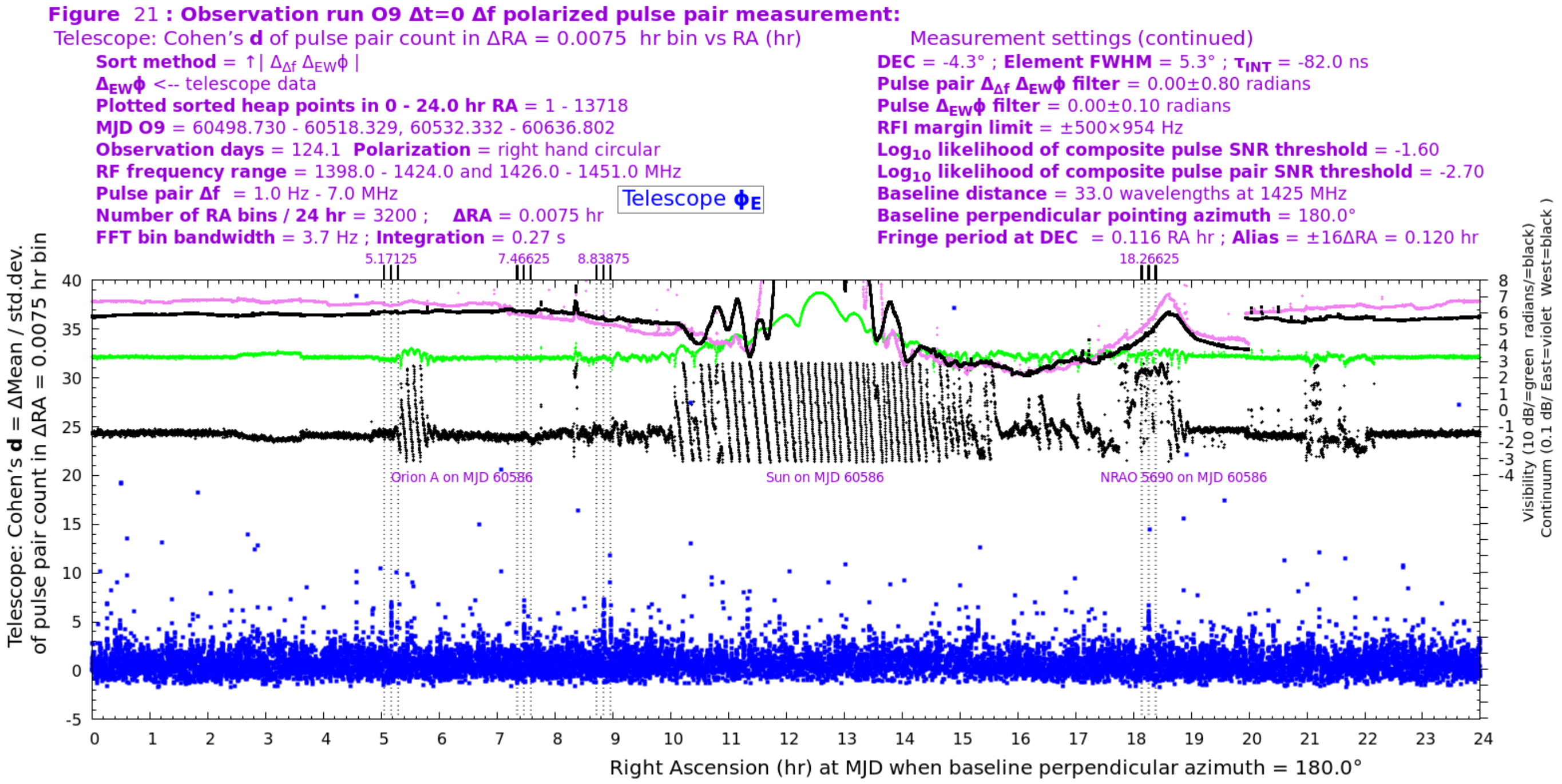

**Figure 21:** Cohen's **d** statistical population effect size, observed during the 124.1 day experiment, is plotted together with MJD 60586 50 MHz bandwidth, 0.27 s integrated, continuum and complex visibility interferometer measurements. Effect size across the RA bin population is evaluated, seeking increasingly high numbers of standard deviations in one RA bin, or two adjacent RA bins, and within RA bins expected to be present due to one wavelength geometric delay interferometer direction aliasing. Four candidate DOIs are studied at higher RA resolution in telescope data, in **Figs. 2-3, 10-18 and 25.**

The absence of the expected sawtooth visibility phase in the NRAO 5690 observation is thought to be due to a slow time-varying broadband noise source in common view of element antenna directivity patterns, and/or within the instrument. If the observed celestial object does not provide high enough antenna temperature to overcome the interferometer common source noise power, then deviations in the visibility phase from the sawtooth shape are expected.

Off-scale solar measurements of east and west element continuum average power are displayed on-scale at reduced scale sensitivity in **Fig. 22.**

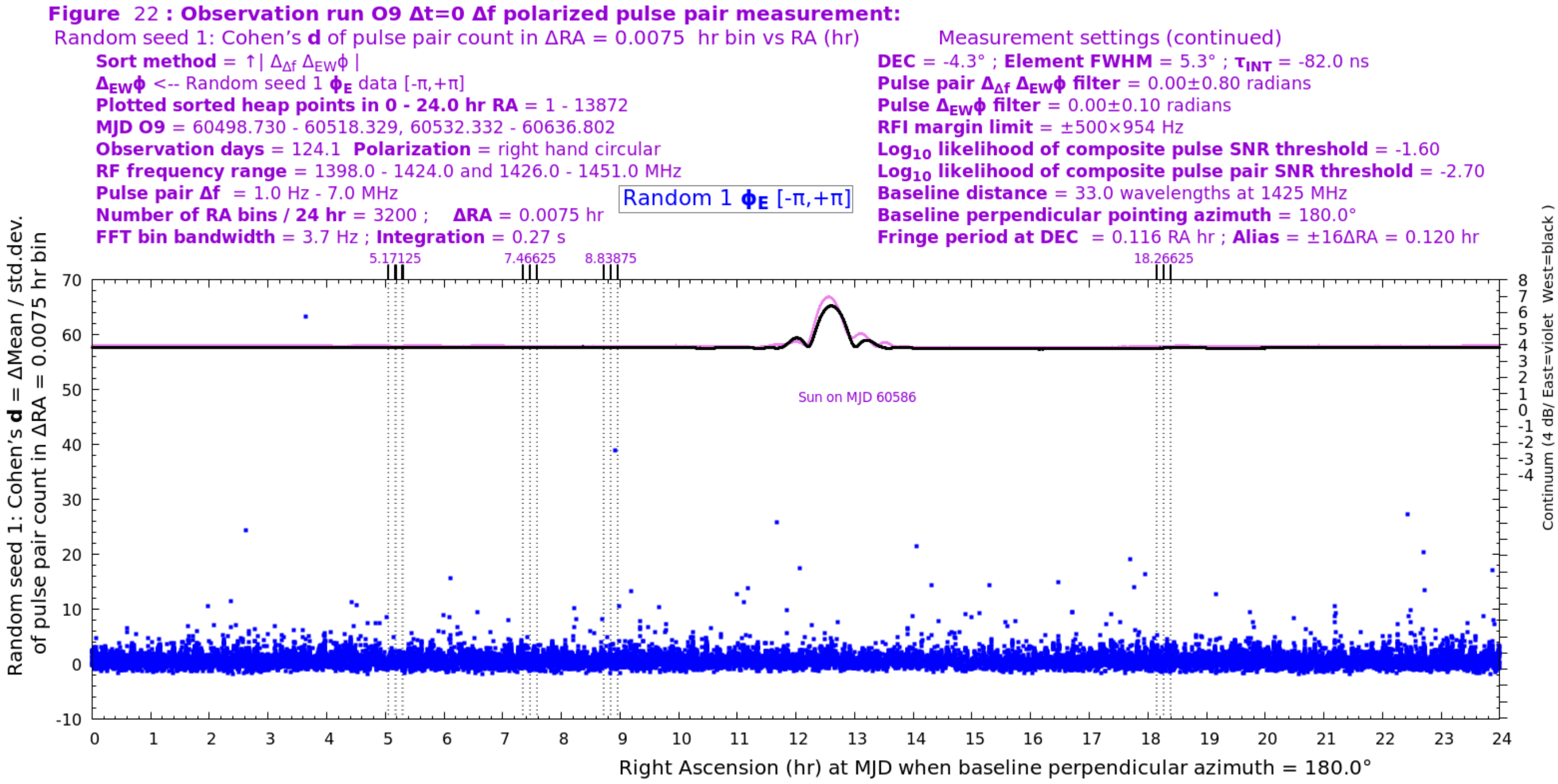

**Figure 22:** Random phase modulation is used in the measurement of Cohen's **d** effect size of pulse pair count. The result is plotted as in **Fig. 21**, with reduced continuum plot sensitivity. The uniform random phase noise replaces the east interferometer element measured phase. The continuum plot is included here to display the off-scale response of **Fig. 21**.

Highest Cohen's **d** values result from pulse pairs present at the top of the sorted heap, due to their low event probability. Comparing this plot to **Fig. 21** shows indications that the anomalous consistently high Cohen's **d** values in the DOIs are affected by the phase measurements of the east interferometer element, indicating a common direction of arrival of simultaneous pulses in each pulse pair.

---

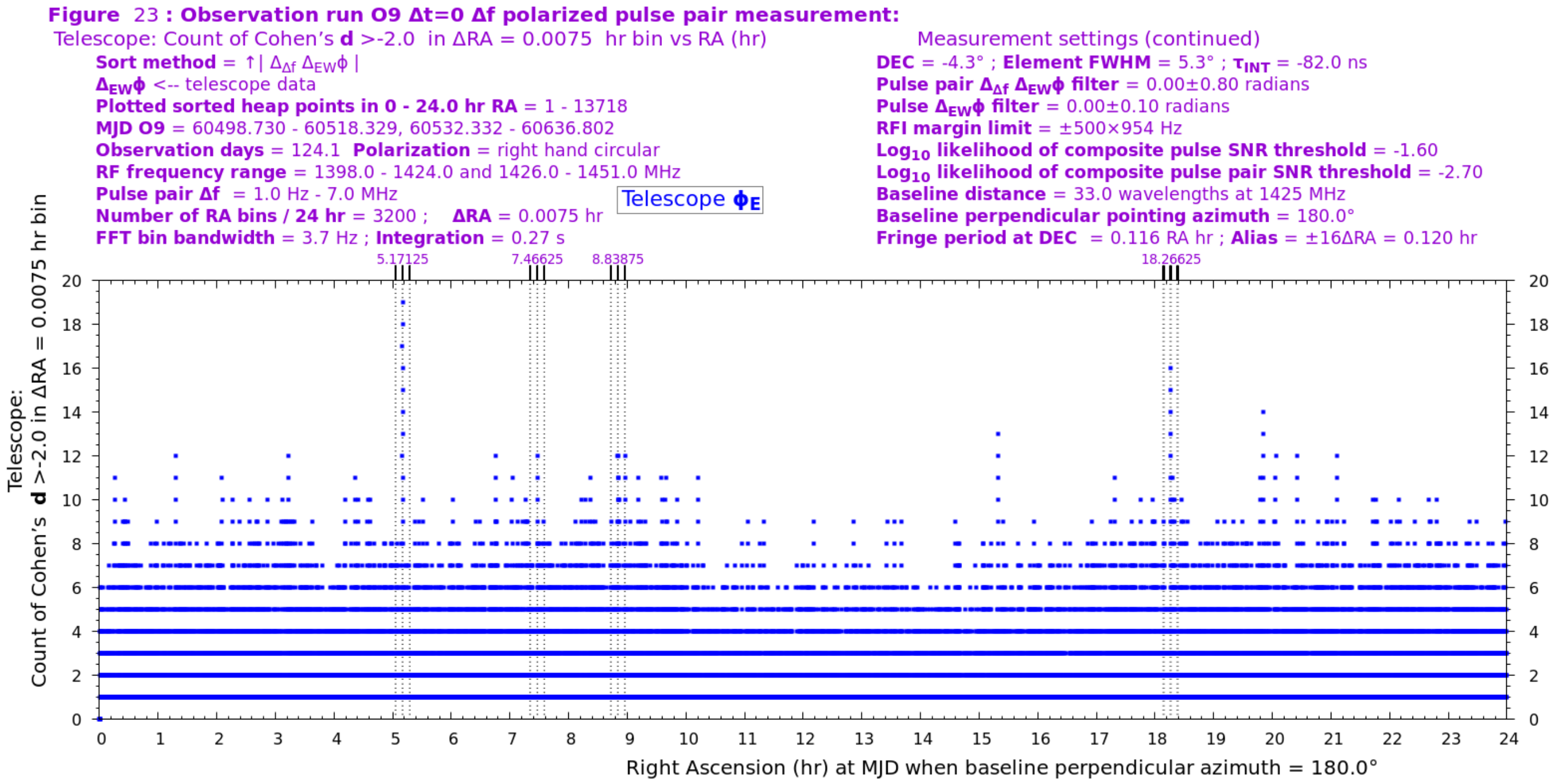

**Figure 23:** The count of pulse pairs in each RA bin is displayed. The 7.46625 hr and 8.83875 hr RA DOIs have a significant number of pulse pair counts in adjacent RA bins, resulting in a total count of 20 and 24 pulse pairs, respectively, apparent in the enhanced RA resolution plots **Figs. 3(b-c)**. **Fig. 24** shows the measured counts per RA bin with random phase noise applied to the east element signal phase measurement. The high count values of the DOIs are somewhat higher than the mean count expected with a uniform distribution of counts across all RA bins, i.e. 4.3 pulse pairs per RA bin.

The number of anomalous pulse pair counts may be reduced if pulse pairs are transmitted using a number of pulse durations, to increase discovery channel capacity and ameliorate the mismatch of transmit pulse occupied bandwidth and receiver filters.

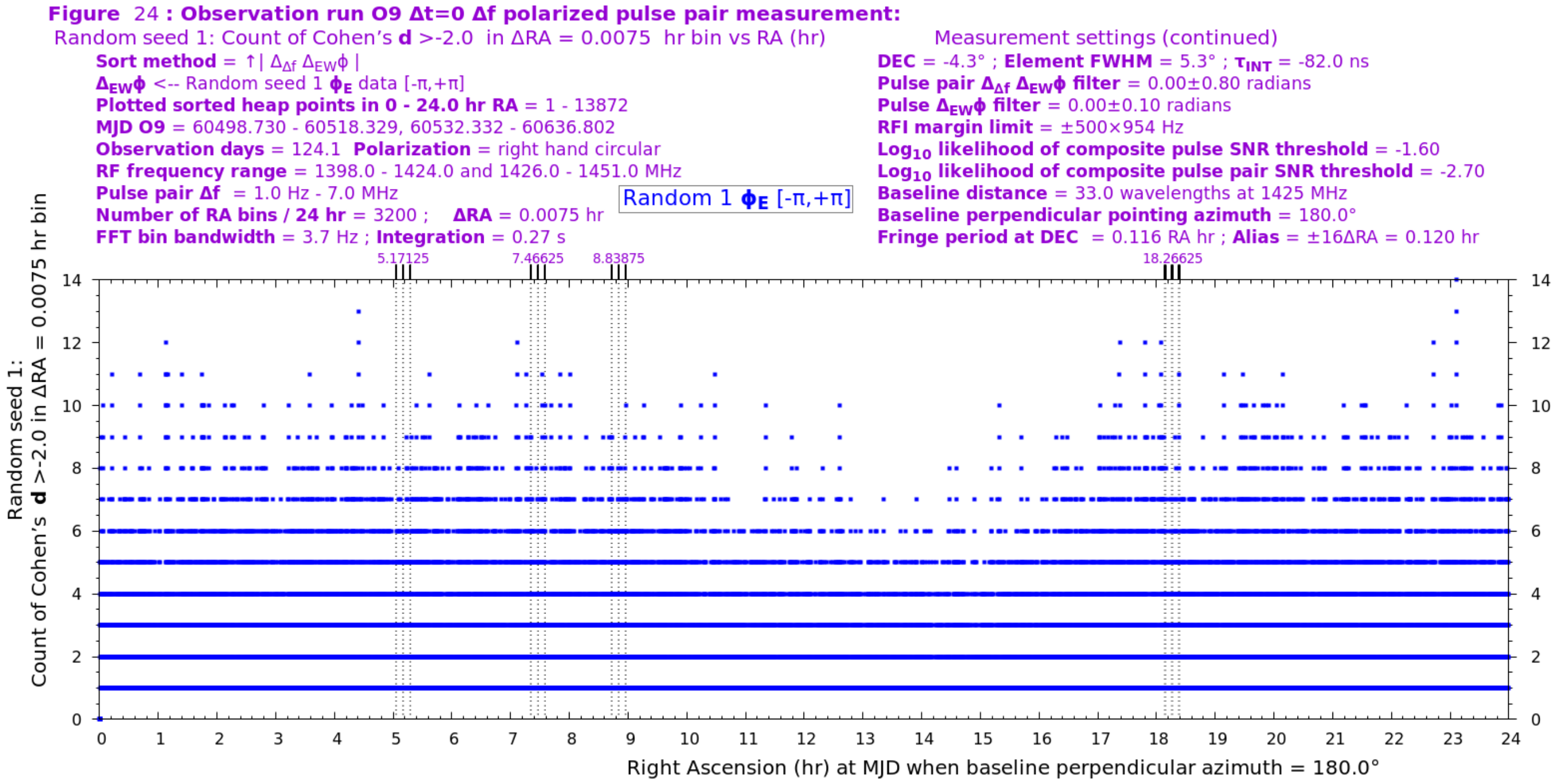

**Figure 24:** Using random phase modulation, the count per RA bin of pulse pairs across the full RA range, provides an estimate of how background and system noise might cause anomalies. Large anomalies do not appear, as observed in **Fig. 23**. Average pulse pair counts appear in this plot higher than the expected value of total pulse pairs / number of RA bins = 4.3 due to the compression of displayed points on the horizontal axis.

---

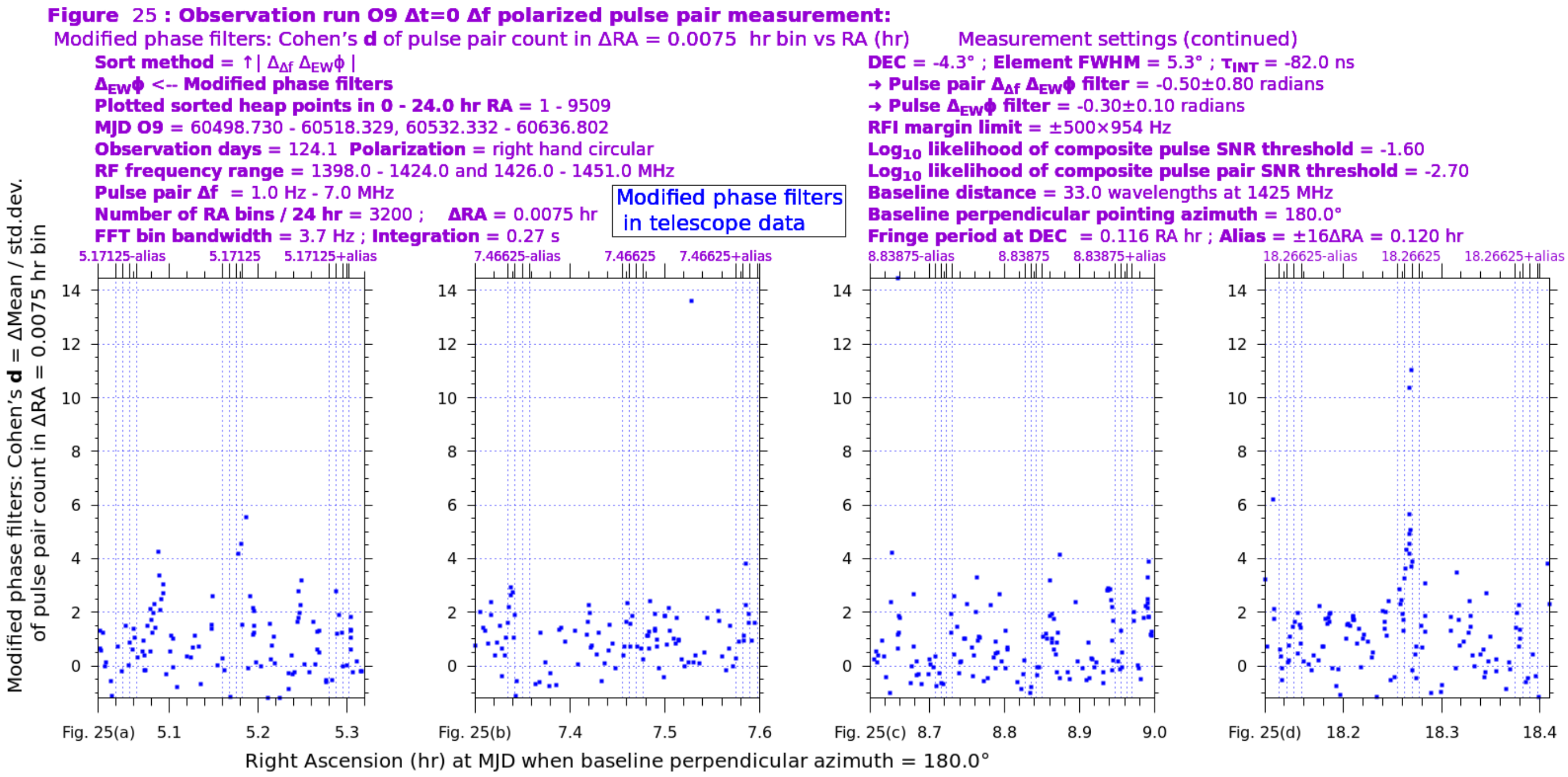

**Figure 25:** Changing the range of the filters of pulse pair phase measurements, $\Delta_{\Delta f}\Delta_{EW}\phi$ and $\Delta_{EW}\phi$, provides a way to examine the phase characteristics of received pulse pairs in a celestial direction. A spatially extended natural celestial object is expected to measure a somewhat random $\Delta_{\Delta f}\Delta_{EW}\phi$, because the high SNR spectral outliers in two simultaneous pulses, Δf spaced, in a pulse pair are expected to have uncorrelated RF phase, pulse pair to pulse pair. On the other hand, a relatively compact-sized aperture of repetitive intentionally transmitted pulse pairs, having components Δf spaced, are expected to have a common angle of arrival, and therefore a consistently low $\Delta_{\Delta f}\Delta_{EW}\phi$ measured value, pulse pair to pulse pair.

The DOI of **Fig. 25(d)** exhibits high values of Cohen's **d** with modified phase filters, implying a wide distribution of phase of the pulses in the pairs. Together with the observation of nine of sixteen pulse pairs in the 18.26625 hr RA bin having a single-day MJD = 60598.89080-60598.89105, partially indicated in **Fig. 18(d)**, and high SNR shown in **Fig. 17(d),** the 18.26625 hr RA DOI may be considered to be unlike the other three DOIs, comparing **Figs. 25(a-d)** to **Figs. 2(a-d)**.

The use of modified filtering of pulse pair signal phase may provide a way to differentiate between natural emissions, RFI and intentional transmissions.

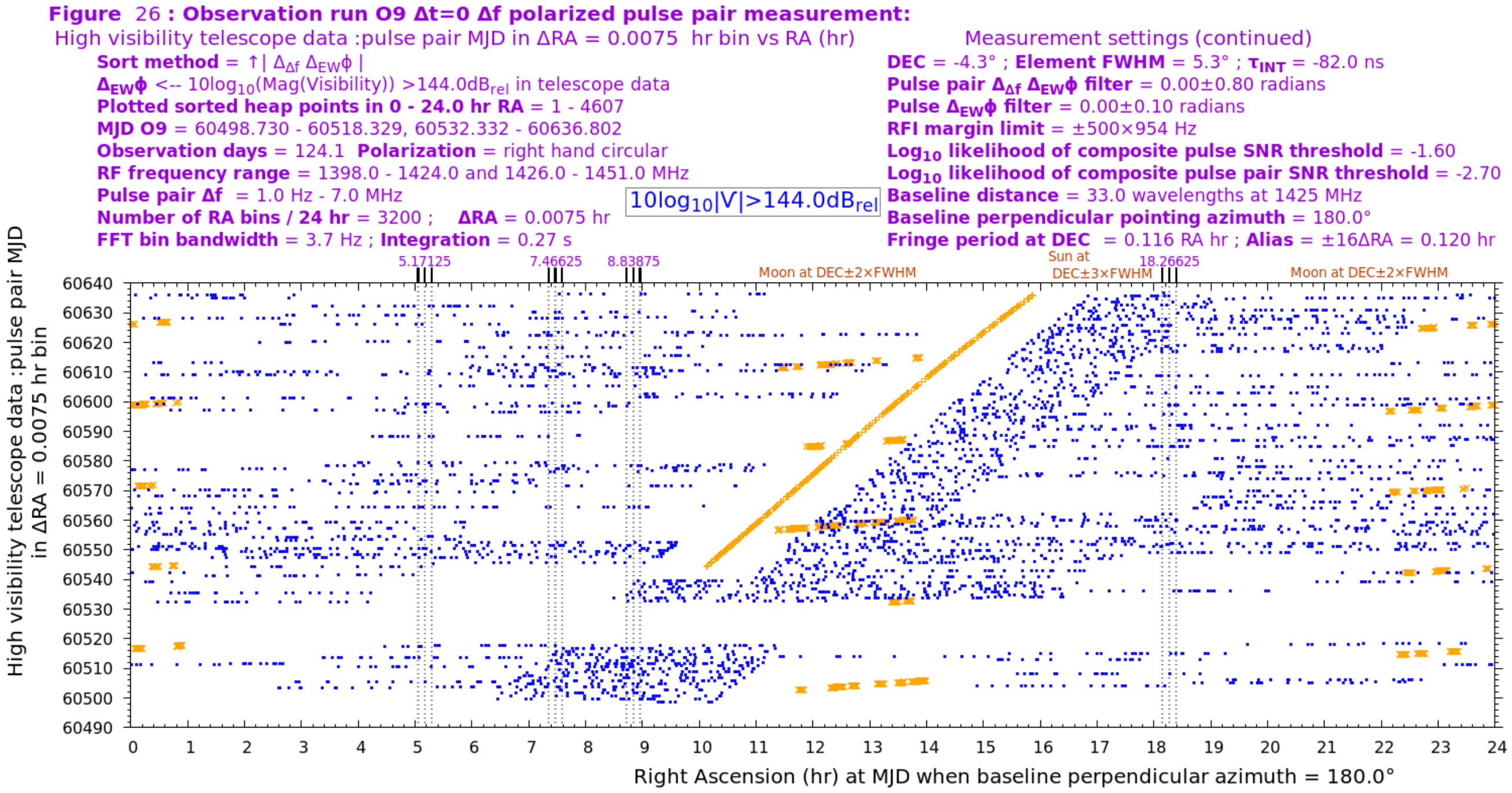

**Figure 26:** Pulse pairs occurring simultaneously with high values of log visibility magnitude, i.e. $10\log_{10} | V | > 144.0$ dBrel, are examined to seek possible causes of high SNR outliers of polarized pulse pairs. The dense concentrations near the Sun-excised region are thought to be due to asymmetric antenna side-lobe response to emissions from the Sun, and possibly reflection from objects local to the antenna elements. Other high visibility outliers appear to be correlated with their MJD day. The discontinuities in the Sun and Moon lines are due to fewer plotted (MJD,RA) data points compared to **Fig. 19**.

## V. Discussion

The difference in pulse pair count observations between telescope phase and modified phase measurements, in **Figs. 4** and **5-9**, respectively, together with the concentration of high Cohen's **d** measurements in DOI RA bins, and concentration in interferometer aliased bins, in **Figs. 2, 3, 4, 21 and 23**, during the 124.1 day experiment, suggest that one or more celestial phenomena seem to be reasonable explanations for the observed anomalies. Further support of a celestial cause is provided by the absence of low value Cohen's **d** measurements in the RA bins of DOIs, as observed in **Fig. 2**. The 5.17125 hr RA DOI has celestial-cause support given prior experiments using different instruments and methods of measurement. **[1][2][3][4]** The 18.26625 hr RA DOI appears different from other DOIs in several associated measurements, e.g. wide band power, Δf, SNR noise-caused likelihood, and MJD, observed in **Figs. 13, 15, 17, 18,** and in the phase filter test results shown in **Fig. 25.** The 18.26625 hr RA DOI may therefore be due to a non-celestial cause.

**Expected amplitude distribution:** Natural astronomical objects generally exhibit Rayleigh distributed amplitude probability distributions, as time and frequency averaging are inherent in receiver signal processing. The contributions of many current elements to the electromagnetic complex vector potential at the telescope antenna feeds determines the expected likelihood of a pair of simultaneous narrow bandwidth pulses each exceeding an SNR threshold. **[9][11][12][13]**

**Classical electromagnetic theory:** Tracing the classical calculation of the propagation of polarized pulse pairs from many transmitting current elements to the receiver antenna feed complex voltages, leads to the understanding that during an interval in time, i.e. the FFT averaging time, within two narrow RF frequency bands, and within a polarization space region on the surface of the Poincare sphere, the many vector potentials at the receiving antenna feed need to add coherently to a high level above the system noise. This restriction applies to natural astronomical objects and to intentional interstellar transmitters.

**Quantum theory:** Individual photons that collectively exhibit the observed pulse pair properties are required to each have nearly identical polarization vector and propagation distance from synchronized energy sourcing current elements, assuming radiation is due to electron acceleration. In addition, photon energies need to be almost equal to be anomalously present in the observed narrow FFT bins. In this experiment,

$$h\nu \ll kT_{SYS}, \qquad (5)$$

where $\nu$ is the narrow bandwidth pulse RF frequency, h Planck's constant, k Boltzmann's constant, and $T_{SYS}$, the receiver system temperature. Photon count per bin measurement is proportional to the product of the FFT bin width, 3.7 Hz, and the FFT integration time 1/3.7 Hz = 0.27s, cancelling. A calculation yields approximately $4\text{x}10^3$ photons contribute to a power level equivalent to a thermal object having 290 K antenna temperature, at each of the two frequencies in the pulse pair. It seems, among natural astronomical objects, that only compact or planar homogeneous, slow-rotating, slow-accelerating multi-resonant MASERs might satisfy this requirement. Such narrow bandwidth pulse pair MASERs seem in theory to be feasible, using Zeeman splitting in non-fluctuating high magnetic fields, in dense clouds of molecules containing atoms in the transition group of elements. **[14]** The stated restrictive characteristics of these types of candidate modeled objects seem to imply reasoning for sparse observational evidence of narrow bandwidth pulse pairs originating from natural phenomena.

**High SNR outliers:** It is possible that a pair of high SNR outlier pulsed signals will exhibit similar pulse pair properties, from time to time. However, such SNR outlier candidate DOIs must collectively exhibit these nearly identical properties from nearly the same celestial pointing direction, over time. It seems likely that such SNR outliers, if due to random processes, would be observed to be distributed over many pointing directions. The observation that frequency spaced simultaneous narrow bandwidth pulses have nearly the same RF phase, and apparently the same celestial pointing direction, seems highly unusual.

**Correlation across antenna elements:** The received signals at the antenna elements are a combination of signals correlated across antenna elements and signals uncorrelated across antenna elements. Wide bandwidth power measurements, together with Cohen's **d** measurements of pulse pair counts and interferometer visibility, may be used to help determine if correlations are present. When correlation is present, expected independent Rayleigh distributions at each element transform into two Ricean distributions, where the in-phase/quadrature (IQ) zero offset inherent in the Ricean distribution model is proportional to the amplitude associated with the cross-correlation of the correlated signals applied to the antenna elements. The increased event probability in the presence of correlated signals is calculated in Appendix B of **[1]** and plotted in Figure 1 of **[3]**. Analysis of wide bandwidth signals coincident with narrow bandwidth pulse pairs is a topic of further work.

**RFI hypotheses:** RFI seems unlikely to be associated with evidence of sustained common celestial source directions during long duration experiments. On the other hand, RFI hypotheses are difficult to falsify because the absence of evidence of an associated RFI source does not contribute to such RFI hypotheses falsifications.

**Impact of a biosignature discovery:** The significance of an interstellar transmitter hypothesis (T) is potentially enhanced by experimental evidence of a galactic biosignature (B), when applied in the Bayesian theorem as follows.

$$Pr(\mathrm{T})\ Pr(\mathrm{B} \mid \mathrm{T}) = Pr(\mathrm{T} \mid \mathrm{B})\ Pr(\mathrm{B})\,, \qquad (6)$$

where $Pr(\bullet)$ is probability and $Pr(\bullet|\bullet)$ is conditional probability. $Pr(\mathrm{B} \mid \mathrm{T})$ may be nearly one, assuming that the signal-based evidence of an interstellar transmitter results from biological processes, while $Pr(\mathrm{T} \mid \mathrm{B})$ may be close to one, assuming that biological-based signal transmitters are long lasting in time. In this simplified analysis, evidence of biosignatures may contribute directly to an inference of an interstellar transmitter, absent auxiliary hypotheses.

**Discoverable communication system design:** A high information capacity communication system need not have signals identical to random noise, assuming that some information capacity is allocated to transmit discoverable signals. Such discoverable signals are expected to have a channel capacity to permit further decoding of the signals. Polarized pulse pairs that cover a wide range of pulse durations, RF frequencies and polarization seem to be a reasonable way to accomplish simultaneous objectives.

## VI. Conclusions

Evidence suggests that observed similar narrow bandwidth polarized pulse pair radio signals are sourced from three celestial directions. Without independent corroboration and sufficient natural object modeling, the cause, or causes, of the pulse pairs reported in this work are unknown.

## VII. Further Work

Construction has begun on a third interferometer element, providing the antenna system with three non-overlapping baselines. The use of the third element allows a 1.5 dB reduction in SNR threshold and the possibility of DEC direction estimation. The study of associated measurements is ongoing. Additional further work is described in previous reports.**[5][6][7][8]** Seeking independent corroboration is important.

## VIII. Acknowledgements

Many friends and my family, during seven decades, have made my SETI work highly enjoyable, and possible. Much gratitude goes to workers in many organizations, including: Green Bank Observatory, Deep Space Exploration Society, National Radio Astronomy Observatory, Society of Amateur Radio Astronomers, SETI Institute, Breakthrough Listen, U.C. Berkeley SETI Research Center, Hewlett-Packard Co., Vivato Inc., Cellular Specialties Inc., Westell Technologies Inc., Penn State Extraterrestrial Intelligence Center, Case Western Reserve University, Spokane Community College, Allen Telescope Array, Pisgah Astronomical Research Institute, HamSCI, SETI League, and equipment and software suppliers.